\documentclass[aps,prl,superscriptaddress,floatfix,nofootinbib,preprintnumbers,eqsecnum,twocolumn,dvipsnames]{revtex4-2}

\counterwithout{equation}{section}
\usepackage{amsmath,float,graphicx,placeins,amssymb,multirow,pgfplots,pgfplotstable}
\usepackage{float}
\usepackage{empheq}
\usepackage{needspace}
\usepackage{tikz}
\usepackage{comment}
\usepackage{enumerate,slashed,cancel,comment,color,bbm,graphicx,rotating,placeins,mathtools,multirow,hhline}
\usepackage[normalem]{ulem}
\usepackage{array}
\usepackage{hyperref}
\usepackage{color}
\usepackage{subcaption}
 \hypersetup
 {
   colorlinks,
   linkcolor={blue!80!black},
   citecolor={blue!70!black},
   urlcolor={blue!70!black}
 }
\usepackage{graphicx}
\usepackage{subcaption}
\usepackage{algorithm}
\usepackage{algpseudocode}
\usepackage{hyperref}
\algtext*{EndFor}

\begin{document}
\sloppy
\raggedbottom

\title{Solving Einstein Field Equations on a Digital Quantum Computer\\
\small{A Quantum Algorithm for Numerical Relativity}}

\begin{abstract}
    In this work, we show how simulations performed on classical computers such as those of Numerical Relativity can be tackled by quantum algorithms for solving systems of partial differential equations. We develop a proof-of-principle quantum algorithm for solving Einstein Field Equationsin the Wahlquist-Estabrook-Buchman-Bardeen(WEBB) tetrad Numerical Relativity formalism, and test it by evolving the Schwarzschild Black Hole spacetime and perturbing it to obtain gravitational Quasinormal Modes. We program the algorithm components for a gate-based, digital quantum computer using the Qiskit software and run it on classical simulators and physical IBM quantum computers through the UKRI National Quantum Computing Centre (NQCC) Quantum Access program and quantify the computational resources and runtime.

\end{abstract}

\author{Clelia Altomonte}
\email[Email address: ]{clelia.altomonte@kcl.ac.uk}

\author{Malcolm Fairbairn}
\email[Email address: ]{malcolm.fairbairn@kcl.ac.uk}

\affiliation{Theoretical Particle Physics and Cosmology, King’s College London,\\ Strand, London WC2R 2LS, United Kingdom}

\maketitle

\setcounter{secnumdepth}{3}


\section{Introduction}

Numerical Relativity is the research field that studies numerical techniques and algorithms to solve the Einstein field equations of relativistic gravitation 
\begin{equation}
G_{\mu \nu}+\Lambda g_{\mu \nu}=\kappa T_{\mu \nu}
\label{efe}
\end{equation}
where $G_{\mu \nu}$ is the Einstein tensor, $\Lambda$ is the cosmological constant, $g_{\mu \nu}$ is the spacetime metric, $\kappa=\frac{8 \pi G}{c^4}$ is a constant where $c$ being the speed of light and $G$ is Newton's gravitational constant, and $T_{\mu \nu}$ is the energy-momentum tensor. Numerical Relativity techniques allow the study of physical systems beyond those where the simplifying assumptions lead to an analytic solution of eq.(\ref{efe}), by translating the Einstein field equations to systems of partial differential equations (PDEs) that are discretised and recursively solved through software and codes that are however computationally expensive, often requiring running on supercomputers and clusters. In this project, we explore the use of quantum computing methods as an alternative computational method to solve the Einstein field equations. \\

Quantum computing is the computational model that follows the rules of quantum mechanical statistics \cite{Nielsen_Chuang_2012}. As such, it unlocks purely quantum mechanical resources such as quantum superposition, interference, and entanglement.  In leveraging these rescources, some quantum algorithms provide a more efficient way to tackle specific tasks with respect to classical counterparts, quantifiable through the computational complexity framework \cite{vaezi2024quantumcomplexityvsclassical} in terms of runtime as a function of input size, which is referred to as a \textit{quantum advantage}. 

While it is important to notice that such theoretical quantum advantage can be hindered, on the practical side, by the equally quantum mechanical phenomena of decoherence which induces errors, measurement-induced collapse of the wavefunction describing the state of the system (i.e. to extract information from the system, quantum tomography or multuple runs are required), and open research problems such as that of a resource-efficient state-preparation, the development of techniques related to error correction, error mitigation and indirect measurement often make it possible to obtain a \textit{quantum utility} \cite{Kim2023-ho} even with the currently available digital Noisy Intermediate Scale Quantum (NISQ) computers\cite{Bharti_2022}.\\

While both analog and digital quantum computation exists, in what follows we concentrate on digital quantum algorithms in a circuit based framework, which can be proven to implement a universal form of computing \cite{Deutsch_1995, Barenco_1995}, since we are interested in solving the problem at hand digitally.

Quantum computing has not yet been applied to the problem of digitally solving systems of PDEs of Numerical Relativity, even if such problem is not dissimilar to solving highly nonlinear PDEs, such as those of fluid dynamics, for which efficient quantum computing methods have recently been developed \cite{Meng_2023,Lewis_2024}. With this project, we propose to close this gap by developing the first digital quantum algorithm for Numerical Relativity, in the hope that the presented proof-of-principle algorithm
will prove to be a useful tool for the Numerical General Relavity and quantum computing communities.

While such quantum algorithms can be described through the circuit and logical gates framework just as classical algorithms, and can implement universal computing through a finite set of gates, it is important to note that quantum gates are reversible (i.e. all operations can be described by unitaries). As a consequence, alternative routes need to be implemented for classical irreversible operations, copying the state of the system (cf. ``no-cloning theorem" of quantum mechanics\cite{Nielsen_Chuang_2012}) or applying non-unitary operators or non-linear functions (cf. arithmetic operations \cite{Ruiz-Perez_Garcia-Escartin_2017,häner2018optimizing,florio2004quantum,kaye2004reversible,Vedral_1996,Takahashi_Kunihiro_2005,Raggi_2020,draper2000addition}). \\

There are now a few well known examples of quantum digital algorithms, famously Shor's factoring algorithm \cite{365700} and Grover's algorithm for unstructured search \cite{coppersmith2002approximate} which respectively achieve an exponential and a quadratic speed up with respect to classical counterparts, while algorithms based on quantum walks \cite{Bepari_2021,Reitzner_2011} have been shown to achieve an up to exponential speed up \cite{Fillion-Gourdeau_Lorin_2018} under certain conditions. 

Some quantum algorithms for the solution of PDEs have been shown to attain up to exponential speed up in run-time and resources with respect to classical counterparts, and have been applied to a variety of PDE problems (cf. \cite{bharadwaj2020quantum,Cao_2013,Pool_Somoza_Lubasch_Horstmann_2022,linden2020quantum,Berry_2014,Fillion-Gourdeau_Lorin_2018}). Depending on the PDEs system, a given quantum computing technique may be more suitable than others. As a few examples, Harrow–Hassidim–Lloyd(HHL)-type algorithms \cite{Harrow_2009} can solve systems of PDEs once they have been mapped onto a system of linear odes (cf. through Carleman linearisation \cite{vaszary2024carlemanlinearizationpartialdifferential} or other mappings \cite{Childs_Liu_Ostrander_2021,an2023theory,Jóczik_Zimborás_Majoros_Kiss_2022}); if the system of PDEs displays some similarity with a quantum Hamilitonian, then the quantum computing technique of Hamiltonian evolution can be applied, where the system of PDEs is mapped onto a Schrodinger-type Hamiltonian (via the so called Schrodingerisation technique \cite{jin2022quantum_a, jin2022quantum_b}) that is then evolved though quantum simulation; finally, hybrid quantum-classical methods exist, like the Variational Quantum Eigensolver\cite{Demirdjian_Gunlycke_Reynolds_Doyle_Tafur_2022,criado2022qade}, which consists of encoding the ansatz of the solution to the system of PDEs in the variational parameters of a quantum circuit which returns a cost function. The variational parameters are then classically optimised.\\

In this work, we make no claims as to the efficacy of our method in providing a quantum advantage over classical simulations, we simply seek to see if the equations of GR can be mapped onto a digital quantum computer framework as an initial proof-of-principle exercise.\\

The paper is organised as follows: in section \ref{identifying}, we identify the Numerical Relativity formalism suitable for the solution of the general relativistic evolution equations through the quantum computing Hamiltonian simulation method; in section \ref{solving}, we detail the mathematical mapping between PDE discretisation to Hamiltonian, and from Hamiltonian to quantum circuit.  Next, in Section \ref{toy model}, we present the physics example that we will use to test the algorithm, and in Section \ref{algorithm} we provide the related quantum circuit components. In Section \ref{running algorithm}, we comment on the results of testing the algorithm both through classical simulators and physical quantum backends, and comment on the time complexity and resources used. Finally in section \ref{conclusion} we conclude by commenting on future improvements and applications. We provide alternative algorithm components implementations and derivations in the Appendix. 

\section{Identifying a suitable Numerical Relativity formalism for quantum computing}
\label{identifying}

In this section, we summarise the Wahlquist-Estabrook-Buchman-Bardeen (WEBB) Tetrad Formulation for Numerical Relativity formalism \cite{Buchman_Bardeen_2003} and compare it to the Arnowitt-Deser-Misner (ADM) \cite{Arnowitt_2008} and other Numerical Relativity formalisms \cite{Baumgarte_Shapiro_2010, Alcubierre_2008} to argue why it is particularly suitable for a proof-of-principle quantum computing implementation.\\

\begin{figure}
    \centering
    \includegraphics[width=\linewidth]{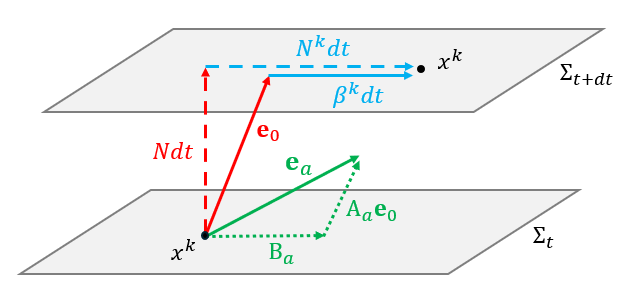}
    \caption{\textit{ Green dotted lines: decomposition of the tetrad leg $\boldsymbol{e}_a$ (green solid line) into a vector tangent to the hypersurface ($B_a$) and a vector parallel to the congruence ($A_a \boldsymbol{e}_0$). 
    Blue solid line: WEBB shift function (=$\beta^k d t$, where $\beta^k$ is the tetrad shift vector) i.e. displacement of the spatial coordinates, $x^k$, relative to the congruence worldline . Blue dashed line: ADM shift function (=$N^k d t$, where $N^k$ is the $3+1$ shift vector) i.e. displacements of $x^k$ relative to the hypersurface normal $n$. Red solid line: WEBB lapse function, proportional to the congruence worldline $\boldsymbol{e}_0$. Red dashed line: ADM lapse function, proportional to the hypersurface normal $n$.}}
    \label{adm webb}
\end{figure}

The ADM formalism for Numerical Relativity is based on a 3+1 foliation of spacetime (3 spatial and 1 time dimensions).

The 3-metric 
$\gamma_{i j}$ and extrinsic curvature $K_{i j}$ live on 3-dimensional hypersurfaces which are updated between timesteps through evolution and constraint equations. A lapse $\alpha$ and $\beta_i$ shift function connect hypersurfaces at different timesteps and constitute a gauge choice (cf. fig.\ref{adm webb}). The 4-metric 
\begin{equation}
g_{a b}=\left(\begin{array}{cc}-\alpha^2+\beta_l \beta^l & \beta_i \\ \beta_j & \gamma_{i j}\end{array}\right)
\end{equation}
that completely determines the spacetime can then be reconstructed at each time by updating the 3-metric and the extrinsic curvature according the the Einstein Field Equations that are decomposed into the the following evolution equations
\begin{equation}
    \partial_t \gamma_{i j}=-2 \alpha K_{i j}+D_i \beta_j+D_j \beta_i
    \label{ADM1}
\end{equation}
\begin{equation}
\begin{aligned}
\partial_t K_{ij} = {} & \alpha\left(R_{ij} - 2 K_{ik} K_j^k + K K_{ij}\right)
- D_i D_j \alpha \\
& - 8\pi \alpha\left(S_{ij} - \frac{1}{2}\gamma_{ij}(S - \rho)\right)
+ \beta^k \partial_k K_{ij}\\
& + K_{ik} \partial_j \beta^k
+ K_{kj} \partial_i \beta^k
\end{aligned}
\label{ADM2}
\end{equation}

where the matter source terms appearing in the above equations are defined by $\rho=n_a n_b T^{a b}, \quad S_{i j}=\gamma_{i a} \gamma_{j b} T^{a b}, \quad S=\gamma^{i j} S_{i j}$\\.  

Eqs.(\ref{ADM1}-\ref{ADM2}) constitute a PDE system that is weakly hyperbolic, first order in time and second order in space. In addition a Hamiltonian and momentum constraint equation are used to find data on the initial hypersurface, and monitor the stability of a given numerical method during the evolution. This constitutes a Cauchy problem (or \textit{initial value problem}), where $\gamma_{i j}$, $K_{i j}$ and source terms are specified on the initial spacelike hypersurface and the evolution equations then determine the future development of the spacetime. 

As they stand, the above coupled PDEs cannot be implemented easily on quantum computers, since they include a large number of operations such as multiplications across a large number of components. While it might be possible to apply Carlemann linearization or Schrodingerisation methods to bring the system to a form of easier quantum computing implementation, it is worth exploring if alternative Numerical Relativity formalisms may already be in such a form.\\

\subsubsection{On the choice of Numerical Relativity formalism used}

There are a number of alternative and often ingenious metric based Numerical Relativity formalisms in the literature such as Generalized Harmonic\cite{Pretorius_2005, Lindblom_2006}, Z4/CCZ4\cite{ali2026solvingsystemslinearequations, Dumbser_2018}, first-order BSSN variants\cite{Bernuzzi_2010}, Einstein-Christoffel formulations \cite{Anderson_1999}.\\

\begin{itemize}
    \item The generalized harmonic gauge formalism introduces gauge source functions $H^\mu=-\Gamma^\mu$ such that Einstein's equations become quasilinear wave equations for the metric
\end{itemize}
\begin{equation}
g^{\alpha \beta} \partial_\alpha \partial_\beta g_{\mu \nu}=S_{\mu \nu}(g, \partial g, H, \partial H) 
\end{equation}
which can be solved either directly in this second-order form, or turned into a first-order system by introducing $\Pi_{\mu \nu}=-n^\alpha \partial_\alpha g_{\mu \nu}$ and $\Phi_{i \mu \nu}=\partial_i g_{\mu \nu}$, such that the evolved statevector is $u=\left(g_{\mu \nu}, \Pi_{\mu \nu}, \Phi_{i \mu \nu}\right)$, and the principal part 
\begin{equation}
\begin{aligned}
&\partial_t g_{\mu \nu} \simeq-\alpha \Pi_{\mu \nu}\\
&\partial_t \Pi_{\mu \nu} \simeq-\alpha g^{i j} \partial_i \Phi_{j \mu \nu}\\
&\partial_t \Phi_{i \mu \nu} \simeq-\alpha \partial_i \Pi_{\mu \nu}
\end{aligned}
\end{equation}
Finally, constraint damping adds terms such as $-\gamma_0 n_{(\mu} C_{\nu)}$, where $C^\mu=H^\mu+\Gamma^\mu$, so that the resulting system is symmetric hyperbolic and linearly degenerate.\\

\begin{itemize}
    \item The Z4 formalism, instead, enlarges Einstein's field equations through a constraint vector $Z_\mu$, i.e.
\end{itemize}
\begin{equation}
R_{\mu \nu}+\nabla_\mu Z_\nu+\nabla_\nu Z_\mu=8 \pi\left(T_{\mu \nu}-\frac{1}{2} g_{\mu \nu} T\right)
\end{equation}
The principal evolution structure becomes 
\begin{equation}
\begin{aligned}
&\partial_t \gamma_{i j}=-2 \alpha K_{i j}\\
&
\partial_t K_{i j} \simeq-\nabla_i \nabla_j \alpha+\alpha \partial_{(i} Z_{j)}
\end{aligned}
\end{equation}
with constraints satisfying wave-type equations
\begin{equation}
\begin{aligned}
&\partial_t \Theta \simeq \partial_i Z^i\\
&\partial_t Z_i \simeq \partial_i \Theta
\end{aligned}
\end{equation}
\begin{itemize}
    \item Standard BSSN uses the variables $\left(\phi, \tilde{\gamma}_{i j}, K, \tilde{A}_{i j}, \tilde{\Gamma}^i\right)$, such that the principal part becomes
\end{itemize}
\begin{equation}
\begin{aligned}
&\partial_t \tilde{\gamma}_{i j}=-2 \alpha \tilde{A}_{i j}\\
&\partial_t \phi=-\frac{1}{6} \alpha K\\
&\partial_t K \simeq-\gamma^{i j} \partial_i \partial_j \alpha\\
&\partial_t \tilde{A}_{i j} \simeq R_{i j}^{T F}\\
&\partial_t \tilde{\Gamma}^i \simeq-2 \partial_j\left(\alpha \tilde{A}^{i j}\right)
\end{aligned}
\end{equation}
A first order reduction introduces other variables such as $d_{k i j}=\partial_k \tilde{\gamma}_{i j}$ and $A_i=\partial_i \ln \alpha$, thus working with an enlarged statevector $(\tilde{\gamma}, \tilde{A}, K, \tilde{\Gamma}, d, A, \ldots)$ and making the system strongly hyperbolic.
\begin{itemize}
    \item Similarly, the conformal covariant (CCZ4) formalism uses the variables $\left(\tilde{\gamma}_{i j}, \tilde{A}_{i j}, \chi, K, \Theta, \hat{\Gamma}^i\right)$, where $\hat{\Gamma}^i=\tilde{\Gamma}^i+2 \tilde{\gamma}^{i j} Z_j$ is the confirmal connection variable. The evolution equations are of the type  
\end{itemize}
\begin{equation}
\begin{aligned}
&\partial_t \chi=\frac{2}{3} \chi\left(\alpha K-\partial_k \beta^k\right)\\
&\partial_t \tilde{\gamma}_{i j}=-2 \alpha \tilde{A}_{i j}\\
&\partial_t K=\cdots-3 \alpha \kappa_1\left(1+\kappa_2\right) \Theta\\
&\partial_t \Theta=\cdots-\alpha \kappa_1\left(2+\kappa_2\right) \Theta\\
&\partial_t \hat{\Gamma}^i=\cdots-2 \alpha \kappa_1 \tilde{\gamma}^{i j} Z_j 
\end{aligned}
\end{equation}
where parameters $\kappa_1, \kappa_2$ provide explicit constraint damping. Hence, it has similarities with both BSSN and the Z4 formalisms due to the propagating constraints and constraint damping. 
\begin{itemize}
    \item The Einstein-Christoffel formulation uses variables $\left(\gamma_{i j}, K_{i j}, f_{k i j}\right)$, where $f_{k i j}=\Gamma_{(i j) k}+$ trace adjustments, so that the principal part becomes 
\end{itemize}
\begin{equation}
\begin{aligned}
&\partial_t \gamma_{i j}=-2 \alpha K_{i j}\\
&\partial_t K_{i j} \simeq-\alpha \partial^k f_{k i j}\\
&\partial_t f_{k i j} \simeq-\alpha \partial_k K_{i j}
\end{aligned}
\end{equation}
which exhibits a first-order wave-system structure in the pair $\left(K_{i j}, f_{k i j}\right)$ and can be demonstrated to be strongly hyperbolic.\\

Notice that, while the above metric-based formulations may be more suitable for a quantum computing implementation since they simplify the evolution equations, they do so at the cost of introducing extra variables, and still require evolving a large number of tensor components and metric derivatives in a non-trivial way, which would need to be discretised and encoded in vector-like structure in such a way as to trying to make the computation efficient. 
\begin{itemize}
    \item Tetrad-based formulations may provide a a more compact alternative way of encoding the data, since the main quantities evolved are connection coefficients, which already store derivatives of the metric, thus putting it on an equal footing with the extrinsic curvature, and, in the case of the WEBB hyperbolic tetrad formalism, producing evolution pairs of the type 
\end{itemize}
\begin{equation}
\begin{aligned}
&\partial_t K_{a b} \simeq-\partial_c N_{a b}\\
&\partial_t N_{a b} \simeq-\partial_c K_{a b}
\end{aligned}
\end{equation}
where $K_{a b}$ are Ricci rotation coefficients, $N_{a b}$ are spatial connecction coefficients, and there are gauge evolution equations for the tetrad acceleration $a_a$ and angular velocity $\omega_a$. The above system is first order hyperbolic symmetrisable, and provides additional advantages for a quantum computing implementation which we consider in detail in the next section.

\subsubsection{WEBB hyperbolic tetrad formalism}

Unlike the ADM formalism, which is a \textit{metric} representation of canonical GR (or a holonomic approach), the WEBB hyperbolic tetrad formalism is \textit{connection-based} and non-holonomic, which means that the fundamental variables evolved are connection coefficients defined over a local, rather than global, coordinate system through frames called tetrads. More specifically, tetrad vector components are orthonormal basis vectors that define local Lorentz frames $\boldsymbol{e}_\alpha(\alpha=0,1,2,3)$. The timelike component, $\boldsymbol{e}_0$, defines the preferred timelike congruence, while $\boldsymbol{e}_1, \boldsymbol{e}_2$, and $\boldsymbol{e}_3$ are spatial triad vectors with a particular rotational orientation with respect to $e_0$. The spacetime metric is Minkowski everywhere: $g_{\alpha \beta}=\boldsymbol{e}_\alpha \cdot \boldsymbol{e}_\beta=\eta_{\alpha \beta}$, with dual basis $e^\alpha$: $\left\langle\boldsymbol{e}^\alpha, \boldsymbol{e}_\beta\right\rangle=\delta_\beta^\alpha$; $\boldsymbol{e}^\alpha \cdot \boldsymbol{e}^\beta=\eta^{\alpha \beta}$. Information about the spacetime curvature is retrieved through the connection coefficients. As an example, for some applications such as gravitational waves, a convenient option is calculating the Weyl tensor components through the Newman-Penrose formalism\cite{Newman_Penrose_1962}, which is also tetrad-based. For 3+1 dimensions, there are 24 distinct connection coefficients (also known as Ricci rotation coefficients) that transform as scalar fields under coordinate transformations and are anti-symmetric in the final two indices:
\begin{equation}
    \Gamma_{\alpha \beta \gamma}=\boldsymbol{e}_\alpha \cdot \nabla_\gamma \boldsymbol{e}_\beta=-\Gamma_{\beta \alpha \gamma}
\end{equation}
with $\nabla$ being the covariant derivative operator with respect to a tetrad component. The tetrad formalism we use as the basis for the quantum computing implementation is the  ``WEBB hyperbolic tetrad formalism" for vacuum gravity \cite{Buchman_Bardeen_2003}, and specifically its adaptation for numerical relativity \cite{Buchman_Bardeen_2005}. The evolution equations within this framework are given by
\begin{equation}
    D_0 \mathbf{q}+M^a D_a \mathbf{q}=\mathbf{S}
    \label{WEBB evolution}
\end{equation}
where $\mathbf{q}$ is a vector including the 24 possible connection coefficients, grouped into two $3 \times 3$ dyadic matrices $N_{a b} \equiv \frac{1}{2} \varepsilon_{b c d} \Gamma_{c d a}$ and $K_{a b} \equiv \Gamma_{b 0 a}$, plus two spatial vectors, the acceleration $a_b=\Gamma_{b 00}$ and the angular velocity (relative to Fermi-Walker transport) $\omega_b=\frac{1}{2} \varepsilon_{b c d} \Gamma_{d c 0}$ of the tetrad frames, eg.
\begin{equation}
\begin{aligned}
& \mathbf{q}=\left(N_{11}, N_{21}, N_{31}, a_1, K_{11}, K_{21}, K_{31}, \omega_1\right. \\
& N_{12}, N_{22}, N_{32}, a_2, K_{12}, K_{22}, K_{32}, \omega_2 \\
& \left.N_{13}, N_{23}, N_{33}, a_3, K_{13}, K_{23}, K_{33}, \omega_3\right)
\end{aligned}
\end{equation}
$M^a$ are sparse, block-diagonal, $24 \times24$ dimensional unitary matrices whose only nonzero elements are $\pm 1$; $D_0$ the derivative with respect to the tetrad congruence; $D_a$ is the spatial derivative with respect to the tetrad spatial components. The structure of the equations in this form is first order symmetrisable hyperbolic (FOSH) with non-linear terms isolated in the $\mathbf{S}$ vector of source terms (the detailed source terms and constraint equations can be found in \cite{Buchman_Bardeen_2003}). Given a choice of coordinates $\boldsymbol{e}_\alpha=\lambda_\alpha^\mu \boldsymbol{e}_\mu$, the components version of the above derivatives become
\begin{equation}
    \boldsymbol{e}_a=A_a \boldsymbol{e}_0+B_a^k \frac{\partial}{\partial x^k}
\end{equation}
\begin{equation}
D_0=\frac{1}{\alpha}\left(\frac{\partial}{\partial t}-\beta^k \frac{\partial}{\partial x^k}\right)
\end{equation}
\begin{equation}
D_a=A_a D_0+B_a^k \frac{\partial}{\partial x^k}
\end{equation}
where the tetrad lapse function $\alpha$ denotes rate of change of proper time with respect to coordinate time along the tetrad congruence while the tetrad shift vector $\beta^k$ denotes rate of displacement of the spatial coordinates relative to the tetrad congruence worldlines per unit coordinate time. It is possible to show that the relation between the ADM lapse function and shift vector is
\begin{equation}
    \alpha_{A D M}= N =\frac{\alpha}{\sqrt{1-A_a A_a}}
\end{equation}
\begin{equation}
    \beta_{ADM}^k= N^k =\beta^k+\frac{\alpha A_a B_a^\kappa}{1-A_a A_a}
\end{equation}
\noindent where $N$ lies along the the normal to the hypersurface $n$.\\

We can now comment on the reasons why this formalism is particularly suited for a quantum computing implementation:

\begin{itemize}
    \item Ricci rotation coefficients transform as scalar fields and  their grid discretization can be compactly amplitude-encoded ($n$ qubits can amplitude-encode up to $2^n$ values) in a vector, an ideal object for quantum computing manipulation, rather than a tensor. 
    \item The evolution equations constitute a simple first-order symmetrisable hyperbolic system: Hamiltonian simulation techniques are easily applicable to such systems since they display similarities with PDE systems for which a quantum computing implementation is known (e.g. \cite{Fillion-Gourdeau_Lorin_2018},\cite{Fillion-Gourdeau_MacLean_Laflamme_2017}), and in some cases is expected to give a quantum advantage with respect to its classical counterpart.  Examples of this are the lattice-Boltzmann and quantum walk techniques \cite{Succi_Fillion-Gourdeau_Palpacelli_2015}, from which some algorithm components have already been developed and can be borrowed.
    \item The evolution equations allow one to combine any non-linear contributions into a single source term $S$, that can be treated with a variety of quantum computing techniques (c.f. \cite{PhysRevE.75.026702,PhysRevE.72.046312,Ollitrault_Miessen_Tavernelli_2021,PhysRevResearch.5.013105,leyton2008quantum,PhysRevResearch.3.023052,Cattaneo_2021,yepez2002efficient,Mezzacapo_2015,Burger_2022}).
\end{itemize}

\begin{figure}
    \centering
    \includegraphics[width=\linewidth]{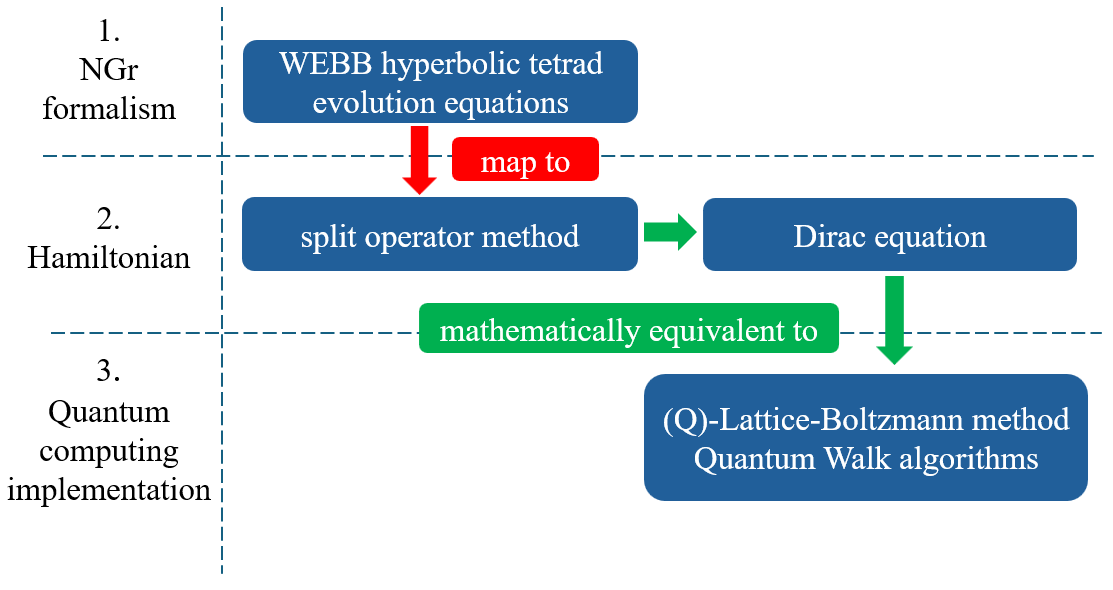}
    \caption{\textit{ Schematic representation of the strategy followed to use quantum computing methods to solve the hyperbolic tetrad evolution equations}}
    \label{strategy}
\end{figure}

Exploiting the points above, we develop a strategy to build a quantum algorithm to solve eq.\ref{WEBB evolution} (cf. fig.\ref{strategy}) which we detail in the following section.

\section{Solving PDEs through Hamiltonian simulation}
\label{solving}

In this section, we present our strategy (cf. fig.\ref{strategy}) to develop a proof-of-principle quantum algorithm for the WEBB Numerical Relativity formalism, and a summary of the relevant techniques and references used.\\

Eq.(\ref{WEBB evolution}) can be solved in its discretised form using a split-operator numerical scheme which can be shown (c.f. \cite{Succi_Fillion-Gourdeau_Palpacelli_2015} and references therein) to be mathematically equivalent to the Quantum Lattice Boltzmann
method \cite{He_Luo_1997, Succi_Benzi_1993} and quantum walk algorithms \cite{Douglas_Wang_2009}  (the quantum analogues of a classical random walk, where the walker is in a superposition of states). The same mathematical mapping has been used in \cite{Fillion-Gourdeau_MacLean_Laflamme_2017} to develop a digital quantum algorithm to solve the Dirac equation (a first-order linear hyperbolic system of PDEs), by exploiting the mathematical equivalence between the lattice discretisation of the Dirac equation in its Majorana form \cite{Fillion-Gourdeau_Herrmann_Mendoza_Palpacelli_Succi_2013} and the split-operator numerical scheme.

In what follows, we provide a brief summary of the methods mentioned above and detail our strategy to solve eq.(\ref{WEBB evolution}) through a quantum walk (quantum walk) by first mapping it to the split-operator method, and use the Dirac equation example to demonstrate this and show how we can borrow the implementation of several operators from the algorithm in \cite{Fillion-Gourdeau_MacLean_Laflamme_2017}. Notice that the mathematical equivalence of the split-operator technique to the Dirac equation Hamiltonian also makes it clear why we choose the digital Hamiltonian simulation technique to solve our system of PDEs, and how it is connected to quantum walk type algorithms.\\

\subsubsection{Split operator method for the Dirac equation}

As shown in \cite{Succi_Fillion-Gourdeau_Palpacelli_2015}, which we follow closely for this summary, the 1-D Dirac equation in Majorana representation (in units where $c=\hbar=1$ )
\begin{equation}
    i \partial_t \psi(z, t)=\left[-i \sigma_z \partial_z+M(z, t)\right] \psi(z, t)
    \label{dirac}
\end{equation}
with the bi-spinor $\psi \in L_2\left(\mathbb{R}, \mathbb{C}^2\right)$, and a generalized mass matrix $M$ which may be used to describe contributions from the physical mass or any kind of coupling, for example to a potential. The solution to eq.(\ref{dirac}), given by
\begin{equation}
    \psi(z, t)=\hat{T} \exp \left[-\Delta t \sigma_z \partial_z-i \int_{t_0}^t M\left(z, t^{\prime}\right) d t^{\prime}\right] \psi\left(z, t_0\right)
\end{equation}
where $\hat{T}$ is the time-ordering operator, $t_0$ is the initial time and $\Delta t=t-t_0$, can be appriximated through an operator splitting technique \cite{Fillion-Gourdeau_Herrmann_Mendoza_Palpacelli_Succi_2013} by
\begin{equation}
\begin{gathered}
\psi(z, t)=\exp \left[-i \Delta t M\left(z, t_0\right)\right] \exp \left[-\Delta t \sigma_z \partial_z\right] \psi\left(z, t_0\right) \\
+O\left(\Delta t^2\right)
\label{split-op}
\end{gathered}
\end{equation}
where the time evolution of the wave function proceeds by a sequence of operations where rotations in spinor space  are followed by space shifts according to
\begin{equation}
\exp \left[-\Delta t \sigma_z \partial_z\right] \psi\left(z, t_0\right)=\left[\begin{array}{l}
\psi_1\left(z-\Delta t, t_0\right) \\
\psi_2\left(z+\Delta t, t_0\right)
\end{array}\right]
\end{equation}
Using a spatial discretization where $\Delta z=\Delta t$ corresponds to a Courant-Friedrichs-Lewy (CFL) condition $C=c \Delta t / \Delta z=1, c$ being the speed of light. Then a single update of eq. (\ref{split-op}) is given by
\begin{equation}
    \left[\begin{array}{l}
\psi_{1, j}^{n+1} \\
\psi_{2, j}^{n+1}
\end{array}\right]=\exp \left[-i \Delta t M_{j, n}\right]\left[\begin{array}{l}
\psi_{1, j-1}^n \\
\psi_{2, j+1}^n
\end{array}\right]
\end{equation}
where $M_{j, n}:=M(n \Delta t, j \Delta z)$ and $\psi_j^n:=\psi(n \Delta t, j \Delta z)$ \\
This numerical scheme is equivalent to a $U(2)$ quantum walk obeying a discrete space-time evolution equation
\begin{equation}
    \left[\begin{array}{l}
\psi_{1, j}^{n+1} \\
\psi_{2, j}^{n+1}
\end{array}\right]=B_{j, n}\left[\begin{array}{l}
\psi_{1, j-1}^n \\
\psi_{2, j+1}^n
\end{array}\right]
\label{generic quantum walk}
\end{equation}
where indices $j, n \in \mathbb{N} \otimes \mathbb{Z}$ label points on a discretisation of space and time, respectively, $B_{j, n}$ is a $U(2)$ operator parametrized by the three space-time dependent Euler angles $\theta_{j, n}, \alpha_{j, n}, \beta_{j, n}$ and a space-time dependent phase $\xi_{j, n}$.
\begin{equation}
    B_{j, n}:=e^{-i \xi_{j, n}}\left[\begin{array}{cc}
e^{i \alpha_{j, n}} \cos \theta_{j, n} & e^{i \beta_{j, n}} \sin \theta_{j, n} \\
-e^{-i \beta_{j, n}} \sin \theta_{j, n} & e^{-i \alpha_{j, n}} \cos \theta_{j, n}
\end{array}\right]
\label{B matrix}
\end{equation}
Notice that this scheme can easily be extended to multiple spatial dimensions (by simply adding $\exp \left[-\Delta t \sigma_x \partial_x\right]$ and $\exp \left[-\Delta t \sigma_y \partial_y\right]$ terms to eq.(\ref{split-op})), or multiple spinor components by substituting the Pauli matrices to rotations in a higher dimensional spinor space (e.g. Gell-Mann matrices, SU(N)) and $B_{j, n}$ with a $U(N)$ operator. \\

\subsubsection{CFL stability conditions}
\label{CFL}

Regarding the stability conditions, notice that (\ref{B matrix}) is exactly unitary for any $U(2)$ matrix $B_{j, n}$. Therefore, viewed purely as a discrete evolution, it is stable regardless of $\Delta t$ and $\Delta x$. Considering instead the continuum limit, we have the standard CFL condition for stability 
\begin{equation}
c_{\mathrm{eff}} \frac{\Delta t}{\Delta x} \leq 1
\end{equation}
and for the quantum walk in eq.(\ref{generic quantum walk}), in dimensionless lattice units we have $c_{\mathrm{eff}} \sim|\cos \theta|$. Therefore changing the angle $\theta$ changes the admissible ratio $\Delta t / \Delta x$ and hence the effective propagation speed ($v_{\max }=|\cos \theta|$), the dispersion relation 
\begin{equation}
\cos (\omega+\xi)=\cos \theta \cos (k-\alpha)
\label{dispersion relation}
\end{equation}
and group velocity 
\begin{equation}
v_g=\frac{\partial \omega}{\partial k}
\end{equation}
which can be seen from calculating $\operatorname{det}\left(U(k)-e^{-i \omega} I\right)=0$ with 
\begin{equation}
U(k)=e^{-i \xi}\left(\begin{array}{cc}
e^{i(\alpha-k)} \cos \theta & e^{i(\beta+k)} \sin \theta \\
-e^{-i(\beta+k)} \sin \theta & e^{-i(\alpha-k)} \cos \theta
\end{array}\right) 
\end{equation}

\subsubsection{Quantum circuit implementation}

In terms of quantum circuit implementations, usually the initial state is amplitude-encoded as a spinor-valued field to reflect the Hilbert space structure of the quantum walk, $\mathcal{H}_{\text {coin }} \otimes \mathcal{H}_{\text {grid }}$,
\begin{equation}
|\Psi\rangle=\sum_{c=0}^n \sum_{x=0}^{m} \psi_c(x)|c\rangle|x\rangle
\label{spinor amplitude encoding}
\end{equation}
where $n$ qubits in the so-called "coin" or "spin" register can encode up to $2^n$ components, and $m\times d$ qubits in the "spatial grid" register encode $2^m$ grid discretisation points for $d$ spatial dimensions. 
Operations can be performed only on the coin register, only on the grid register, or on both. As an example, in the quantum algorithm for solving the Dirac equation in  \cite{Fillion-Gourdeau_MacLean_Laflamme_2017}, the PDE
\begin{equation}
i \partial_t \psi(t, \mathbf{x})=\hat{H} \psi(t, \mathbf{x})
\label{dirac1}
\end{equation}
is solved through the operator splitting method by first decomposing the Hamiltonian as a sum of operators $\hat{H}(t)=\sum_{j=1}^{N_{\text {op }}} \hat{H}_j(t)$. Then the evolution operator
\begin{equation}
\begin{aligned}
\psi\left(t_{n+1}\right) & =T \exp \left[-i \int_{t_n}^{t_{n+1}} \hat{H}(t) d t\right] \psi\left(t_n\right) \\
& =e^{-i \Delta t\left(H\left(t_n\right)+\mathcal{T}\right)} \psi\left(t_n\right)
\end{aligned}
\end{equation}
can be approximated though a Suzuki-Trotter step \cite{SUZUKI_1993}, i.e. as a sequence of exponentials in the form
\begin{equation}
\begin{aligned}
\psi\left(t_{n+1}\right) & =\prod_{k=1}^{N_{\mathrm{seq}}}\left[e^{-i s_0^{(k)} \Delta t \mathcal{T}} \prod_{j=1}^{N_{\mathrm{op}}} e^{-i s_j^{(k)} \Delta t \hat{H}_j\left(t_n\right)}\right] \psi\left(t_n\right) \\
& +O\left(\Delta t^q\right)
\end{aligned}
\end{equation}
Following closely the logic and notation of \cite{Fillion-Gourdeau_MacLean_Laflamme_2017,SUZUKI_1993}, the coefficients $N_{\text {seq }} \in \mathbb{N}^{+}$and $s_j^{(k)} \in \mathbb{R}$ are chosen to get an approximation with a given order of accuracy $q \in \mathbb{N}^{+}$. When some pairs of Hamiltonian in $\left(\hat{H}_i\right)_{i=1, \cdots, N_{\text {op }}}$ do not commute, the splitting induces a numerical error $O\left(\Delta t^q\right)$, where the error can be decreased by changing the value of $q$.

More specifically, for the example considered in \cite{Fillion-Gourdeau_MacLean_Laflamme_2017},
\begin{equation}
    \hat{H}=\boldsymbol{\alpha} \cdot[c \hat{\mathbf{p}}-e \mathbf{A}(t)]+\beta m c^2+e \mathbb{I}_4 V(\mathbf{x}, t)
\end{equation}
where $e$ is the electric charge, $\hat{\mathbf{p}}=-i \boldsymbol{\nabla}$ is the momentum operator, $\mathbf{A}(t)$ the electromagnetic vector potential, while $V(\mathbf{x}, t)=A_0(\mathbf{x}, t)$ is the scalar potential. Through the operator-splitting method, a single update of eq.(\ref{dirac}) is given by
\begin{equation}
\begin{aligned}
\psi\left(t_{n+1}\right) & =Q_{\mathbf{A}}\left(t_n, \Delta t\right) Q_V\left(t_n, \Delta t\right) Q_m(\Delta t)\left[S_z T_z(\Delta t) S_z^{-1}\right] \\
& \times\left[S_y T_y(\Delta t) S_y^{-1}\right]\left[S_x T_x(\Delta t) S_x^{-1}\right] \psi\left(t_n\right)
\end{aligned}
\end{equation}
In terms of quantum circuit components, $\left.Q_a(\Delta t)\right|_{a=x, y, z}=e^{-c \Delta t \alpha_a \partial_a}$ is implemented as $Q_a=S_aT_a S_a^{\dagger}$, where unitary operators $S_a$ (where $a=x, y, z$) transform the Dirac matrices $\alpha_a$ to a Majorana-like representation through $\tilde{\alpha}_a=S_a^{\dagger} \alpha_a S_a=\beta$ which is diagonal with eigenvalues $\pm 1$. This prescribes shifts in opposite directions, implemented on the spatial grid register through a shift operator $T_a(\Delta t)=e^{-c \Delta t \beta \partial_a}$ controlled on the coin register.
The mass term $Q_m(\Delta t):=e^{-i \Delta t \beta m c^2}$ and  vector potential $Q_{\mathbf{A}}(t, \Delta t):=e^{i e \Delta t \boldsymbol{\alpha} \cdot \mathbf{A}(t)}$ operators are applied to the coin register alone due to their global nature, while the scalar potential operator $Q_V(t, \Delta t):=e^{-i \Delta t\left(e \mathbb{I}_4 V(\mathbf{x}, t)\right)}$ is applied to the spatial grid register due to the local dependence on $x$ of $V(\mathbf{x}, t)$.

\subsubsection{Our strategy}

Hence, our strategy to solve eq.(\ref{WEBB evolution}) by means of quantum Hamiltonian simulation, is to map its operator-split discretisation to a quantum walk (i.e. to a Dirac-type Hamiltonian), through

$$
\begin{gathered}
\boldsymbol{q} \rightarrow \psi(t, \mathbf{x}) \rightarrow \text { Amplitude encoding } \\
M^a D_a \rightarrow \boldsymbol{\alpha} \cdot \hat{\mathbf{p}} \rightarrow \text { Advection } \\
\boldsymbol{S} \rightarrow e_4 V(\mathbf{x}, t) \rightarrow \text { Collision }
\end{gathered}
$$

where we conjecture that $S$ can be modelled in the same way as a potential or mass term. This mapping proves to be particularly useful for the problem at hand for the following reasons: 
\begin{itemize}
    \item It allows us to use the operator-split method and to borrow the implementation of some operations from \cite{Fillion-Gourdeau_MacLean_Laflamme_2017} and directly use them in our algorithm. 
    \item It allows one to easily extend the 1-D example to an arbitrary number of dimensions and fields. 
    \item The Hamiltonian simulation and operator splitting literature offer many examples of versatile methods to treat potentials with non-linear terms which we can use to model $S$.
    A few examples include modelling $S$ as a mass matrix, as a collision matrix, or through ancilla-assisted operations on an enlarged Hilbert space\cite{PhysRevE.75.026702,PhysRevE.72.046312,Ollitrault_Miessen_Tavernelli_2021,PhysRevResearch.5.013105,leyton2008quantum,PhysRevResearch.3.023052,Cattaneo_2021,yepez2002efficient,Mezzacapo_2015,Burger_2022}.
    \end{itemize}

In the following section, we introduce the physics toy model that we will use to develop the proof-of-principle quantum algorithm.

\section{Physics toy model: the Schwarzschild black hole solution and its perturbation}
\label{toy model}

We use the example of the Schwarzschild black hole solution within the WEBB hyperbolic tetrad formalism in \cite{Buchman_Bardeen_2005} as basis to construct and test our quantum algorithm. As a way to test the algorithm, we perturb the Schwarzschild solution to produce quasinormal modes (resonant frequencies of a perturbed black hole space-time) following the theory of perturbations of a spherically symmetric spacetime background \cite{Sarbach_2001,Buchman_2007}. In this section, we provide a summary of the relevant equations. 

\subsubsection{Schwarzschild black hole in 1+1 d from ref. \cite{Buchman_2005}}

The spacetime metric is obtained by first calculating $g^{\mu \nu}=\eta^{\alpha \beta} e_\alpha{ }^\mu e_\beta{ }^\nu$. The $g^{\mu \nu}$ matrix is then inverted to give $g_{\mu \nu}$. The resulting metric is
\begin{equation}
\begin{aligned}
& d s^2=\left[-\alpha^2+\beta^{r 2} e^{2 \lambda}\left(1-A_{\hat{r}}^2\right)+2 e^\lambda \alpha \beta^r A_{\hat{r}}\right] d t^2 \\
& \quad+2 e^\lambda\left[\alpha A_{\hat{r}}+\beta^r e^\lambda\left(1-A_{\hat{r}}^2\right)\right] d r d t \\
& \quad+e^{2 \lambda}\left(1-A_{\hat{r}}^2\right) d r^2+R^2 d \theta^2+R^2 \sin ^2 \theta d \phi^2
\end{aligned}
\label{WEBB metric}
\end{equation}
The spherical symmetry of the system allows one to simplify the problem. By using the spherical triad components 
\begin{equation}
\boldsymbol{e}_{\hat{r}}=A_{\hat{r}} \boldsymbol{e}_0+B_{\hat{r}}^r \partial_r, \quad \boldsymbol{e}_{\hat{\theta}}=B_{\hat{\theta}}^\theta \partial_\theta, \quad \boldsymbol{e}_{\hat{\phi}}=B_{\hat{\phi}}^\phi \partial_\phi
\end{equation}
where $r$ is the radial coordinate in the hypersurface, $
B_{\hat{r}}^r \equiv B_R=e^{-\lambda}, \quad B_{\hat{\theta}}^\theta \equiv B_T=\frac{1}{R}, \quad B_{\hat{\phi}}^\phi=\frac{B_T}{\sin \theta},
$ with $R$ being the circumferential radius of the two-sphere. 
As a consequence of the underlying spherical symmetry, the only non-zero Ricci rotation coefficients, which constitute the variables to be evolved, are 
\begin{equation}
K_{\hat{r} \hat{r}} \equiv K_R, \quad K_{\hat{\theta} \hat{\theta}}=K_{\hat{\phi} \hat{\phi}} \equiv K_T, \quad n_{\hat{r}}, \quad a_{\hat{r}}
\end{equation}
where $\boldsymbol{n}=n_{\hat{r}} \boldsymbol{e}_{\hat{r}}$, with Cartesian coordinate components $n_a \equiv \frac{1}{2} \varepsilon_{a b c} N_{b c}$ being the antisymmetric part of $N_{a b}$, which is the only surviving part due to the spherical symmetry.
The Nester gauge\cite{Nester_1992}, which defines special orthonormal frames which make the metric conformally flat.

The evolution equations are given by 
\begin{equation}
D_0 \mathbf{q}+\boldsymbol{C}^{\hat{r}} B_R \partial_r \mathbf{q}=\mathbf{S}
\label{evol schw}
\end{equation}
where 
\begin{equation}
D_0=\frac{1}{\alpha}\left(\partial_t-\beta^r \partial_r\right)
\end{equation}
\noindent and $\alpha$ and $\beta^r$ are the lapse and shift respectively; $B_R$ sets the initial relationship of coordinate radius to proper radius. The vector $q$ encodes the the Ricci rotatio coefficients to be evolved as follows
\begin{equation}
\mathbf{q}=\left(\begin{array}{c}
K_R \\
a_{\hat{r}} \\
K_T \\
n_{\hat{r}}
\end{array}\right)
\label{fields vector}
\end{equation}
where each of $\{K_R, a_r, K_T, n_r\}$ is itself a column vector of $n$ values which are the discretisation of the related Ricci rotation coefficient on $n$ gridpoints. Considering the matrix $C^{\hat{r}}$
\begin{equation}
\boldsymbol{C}^{\hat{r}}=-\frac{1}{1-A_{\hat{r}}^2}\left(\begin{array}{cccc}
A_{\hat{r}} & 1 & 0 & 0 \\
1 & A_{\hat{r}} & 0 & 0 \\
0 & 0 & A_{\hat{r}} & 1 \\
0 & 0 & 1 & A_{\hat{r}}
\end{array}\right)
\end{equation}
a convenient gauge choice consisits of setting $A_{\hat{r}}$ to 0 at the beginning and make sure it remains small by implementing the appropriate lapse and shift evolution or reset operations (this is a hypersurface-orthogonal or zero tilt gauge). This gauge choice makes $C^{\hat{r}}$ a unitary matrix. In what follows we adopt this gauge choice solely to simplify the proof-of-principle algorithm implementation - it does not constitute a fundamental restriction to the framework. In Appendix \ref{appendix implementing} we comment on how to implement a $A_{\hat{r}} \neq 0$ gauge choice in the quantum algorithm. \textbf{S} is a vector of so-called ``source" terms
\begin{equation}
\mathbf{S}=\frac{1}{1-A_{\hat{r}}^2}\left(\begin{array}{c}
S_{-} K_R+A_{\hat{r}} S_{-} a_{\hat{r}} \\
S \_a_{\hat{r}}+A_{\hat{r}} S_{-} K_R \\
S \_K_T+A_{\hat{r}} S_{\_} n_{\hat{r}} \\
S \_n_{\hat{r}}+A_{\hat{r}} S_{-} K_T
\end{array}\right)
\end{equation}
where 
\begin{equation}
\begin{array}{l}
S \_K_R=a_{\hat{r}}^2-n_{\hat{r}}^2-K_R^2+K_T^2+\frac{2 n_{\hat{r}}}{R}, \\
S \_K_T=\frac{a_{\hat{r}}+n_{\hat{r}}}{R}+K_R K_T-K_T^2-n_{\hat{r}}^2-a_{\hat{r}} n_{\hat{r}}, \\
S \_n_{\hat{r}}=\frac{K_T-K_R}{R}-a_{\hat{r}} K_T+n_{\hat{r}}\left(K_R-2 K_T\right) \\
S_{-} a_{\hat{r}}=\frac{2\left(K_R-K_T\right)}{R}
\end{array}
\end{equation}.\\

Additional constraint equations that are used to solve the inital value problem of finding suitable initial values for $A_{\hat{r}}, K_T, n_{\hat{r}}, B_R, K_R$, and $a_{\hat{r}}$ with which to begin the numerical evolution.\\

Finally, there are additional evolution and constraint equations for triad vector components, but we won't need these because $B_R$ can be fixed at the beginning without loss of generality, $A_{\hat{r}}$ is assumed to be 0 at the beginning, and remains small if the following lapse and shift evolution (or periodic resetting) are implemented: 
\begin{equation}
B_R \partial_r(\ln \alpha)=a_{\hat{r}}
\label{lapse function}
\end{equation}
\begin{equation}
B_R \partial_r\left(\beta^r / B_R\right)=-\alpha K_R
\end{equation}
With the above choice of lapse and shift, the relationship of Schwarzschild time coordinate ($d t_{S}$) to the coordinate time used in the simulation ($d t_{s}$) is given by 
\begin{equation}
d t_{S}=\frac{\alpha}{1-R n_{\hat{r}}} d t_{s}
\end{equation}
The \textit{momentum} and \textit{energy constraint equations} used to calculate initial conditions and check the accuracy of the numerical method are respectively 
\begin{equation}
\begin{aligned}
& B_R \partial_r K_T=\left(K_R-K_T\right)\left(\frac{1}{R}-n_{\hat{r}}\right) \\
& \quad-\frac{A_{\hat{r}}}{2}\left[\frac{2}{R}\left(a_{\hat{r}}-n_{\hat{r}}\right)-3 K_T^2-2 a_{\hat{r}} n_{\hat{r}}+n_{\hat{r}}^2\right]
\end{aligned}
\label{momentum constraint}
\end{equation}
\begin{equation}
\begin{aligned}
& B_R \partial_r n_{\hat{r}}=-\frac{2 n_{\hat{r}}}{R}-K_R K_T-\frac{K_T^2}{2}+\frac{3 n_{\hat{r}}^2}{2} \\
& \quad+A_{\hat{r}} K_T\left(a_{\hat{r}}+n_{\hat{r}}\right) 
\end{aligned}
\label{energy constraint}
\end{equation}
Finally, notice that there are also evolution equations for the triad vector components $B_R, B_T, A_r$, and a constraint equation for $B_T$.

\subsubsection{Quasinormal modes}

\begin{figure*}
    \centering
    \begin{minipage}{0.6\textwidth}
        \centering
        \includegraphics[width=\linewidth]{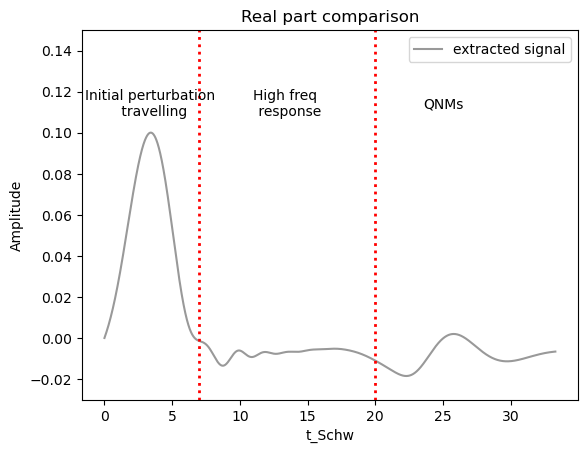}
        \caption{\textcolor{blue}{\textit{ Re$(\Psi_4)$ extracted from the classical finite difference simulation. The signal can be divided into three parts: i) the initial perturbation travelling towards the horizon, ii) the high frequency response, iii) the QNMs ringdown}}}
    \end{minipage}
    \hspace{1cm}
    \begin{minipage}{0.65\textwidth}
        \centering
        \hspace*{0.4cm}
        \includegraphics[width=\linewidth]{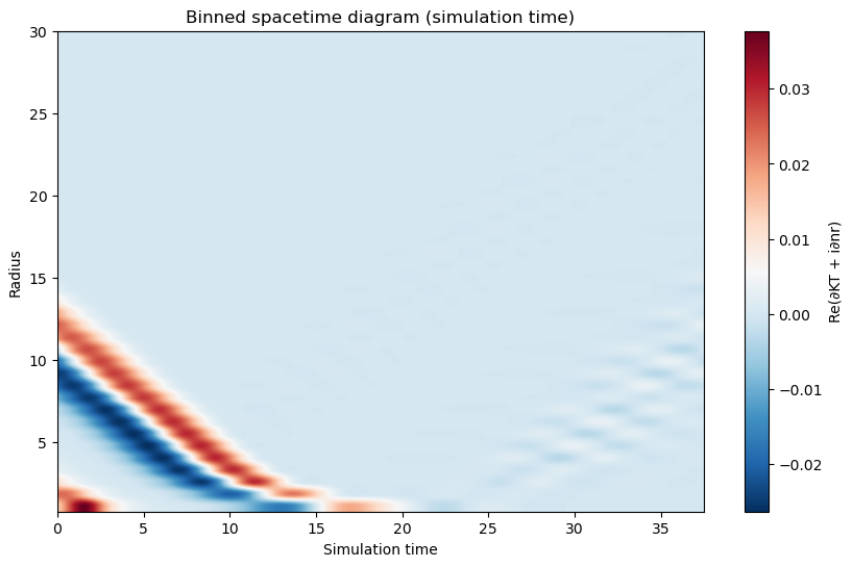}
        \caption{\textit{Spacetime diagram showing the signal over all of the radial bins. Notice the perturbation travelling towards the horison and the QNMs travelling outwards from the horizon following lightcones.}}
    \end{minipage}
\end{figure*}

Quasinormal modes (QNMs) are the resonant frequencies of a perturbed black hole\cite{Sberna_2022, Pazos_2010, Kokkotas_1999}, with frequencies and damping determined solely by the black hole parameters. The Regge-Wheeler ($\Phi_{\ell}^m$) and Zerilli ($\Psi_{\ell}^m$) functions govern the odd and even parity sectors of metric perturbations respectively, and obey
\begin{equation}
\begin{aligned}
& \square \Phi_{\ell}^m-V_{\mathrm{RW}} \Phi_{\ell}^m=0 \\
& \square \Psi_{\ell}^m-V_{\mathrm{Z}} \Psi_{\ell}^m=0
\end{aligned}
\end{equation}
where $V_{\mathrm{RW}}$ and $V_{\mathrm{Z}}$ are the Regge-Wheeler and Zerilli potential respectively. Spin-2 quasinormal modes can be expanded in tensor spherical harmonics as 
\begin{equation}
\Phi_{\ell}^m, \Psi_{\ell}^m = \chi(t, r, \theta, \phi)=\sum_{\ell m} \frac{\chi_{\ell m}(r, t)}{r} Y_{\ell m}(\theta, \phi)
\end{equation}
The frequencies can be fitted from the ringdown segment of a gravitational wave signal, which can be extracted through the Weyl scalar $\Psi_4$, with function of the form 
\begin{equation}
f(t)=A e^{b t} \sin \left[a\left(t-t_0\right)\right]
\end{equation}
As per Birkhoff's theorem \cite{Ghosh_2025}, an exactly spherically symmetric spacetime cannot support gravitational radiation. However, in Appendix section \ref{appendix qnms}, we provide the details as to why a perturbation of $K_T$ and $n_r$ can be projected onto the spin-2 perturbation sector and modulate the QNMs frequencies through the presence of coupling terms in $\Psi_4$ between perturbations of the background metric and the metric components directly responsible for the radiative degrees of freedom. In particular, we find  
\begin{equation}
    \Psi_4 \propto \partial_{t}\delta K_T -i\partial_{t}\delta n_r
    \label{Psi 4 qnms approx}
\end{equation}
and use the frequencies extracted from this function to test the proof-of-concept quantum algorithm against classical solutions of the PDE system (eq. \ref{WEBB evolution}).


\section{Algorithm components}
\label{algorithm}

In this section, we show how to code the algorithm components. For each algorithm component, we present a summary of the high-level details in pseudo-code format, then a detailed description of the mathematical details in the main text, and details of the Qiskit coding implementation in the figures. In this section, we include only the quantum circuit components that we tested on classical simulator and quantum backends, while we include alternative and more general constructions for future improvements in the Appendix sections \ref{appendix implementing},\ref{appendix collision}, \ref{appendix dynamical lapse}, \ref{appendix boundary}.

\subsection{State preparation}

\begin{figure*}[ht]
\centering

\begin{subfigure}[t]{0.45\textwidth}
    \hspace*{0.3cm}
    \centering
    \includegraphics[width=\linewidth]{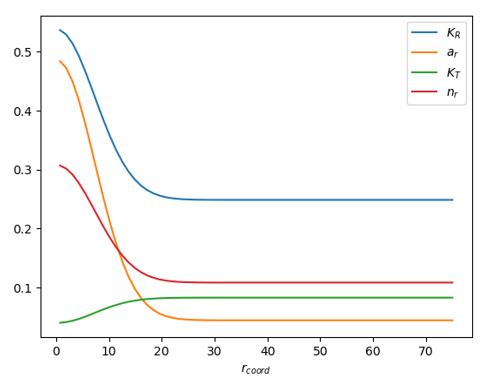}
    \caption{\textit{ Radial grid discretisation of the Ricci rotation coefficients in eq. (\ref{fields vector})}}
    \label{init pert _1}
\end{subfigure}
\hfill
\begin{subfigure}[t]{0.47\textwidth}

    \centering
    \raisebox{0.1cm}{%
    \includegraphics[width=\linewidth]{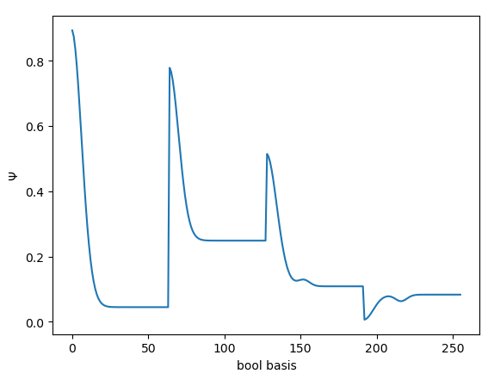}
    }
    
    \caption{\textit{head-to-tail encoding of the fields $[a_r,K_r,n_r,K_T]$, including gaussian perturbations of the form of eq.\ref{perturb} in $n_r$ and $K_T$.}}
\end{subfigure}

\vspace{0.5cm}

\begin{subfigure}[t]{0.5\textwidth}
    \centering
    \includegraphics[width=\linewidth]{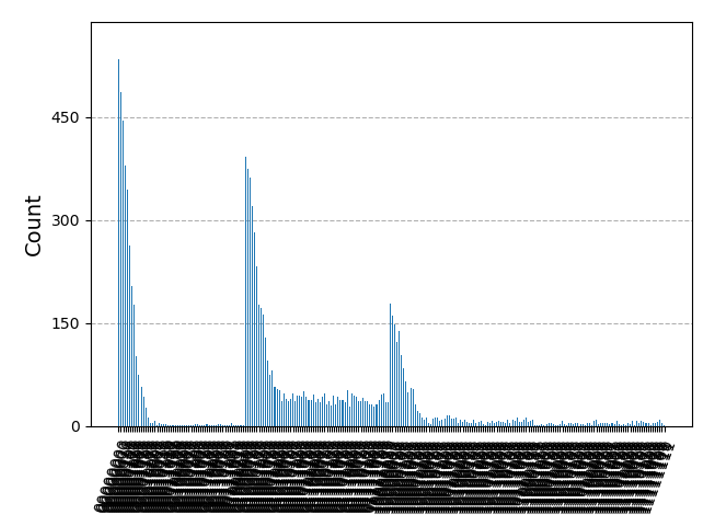}
    \caption{\textit{ Measurement of initialised state using aer-simulator. Notice the little-endian convention: qubits are initialised and measured following increasing coefficient number.}}
\end{subfigure}
\hfill
\begin{subfigure}[t]{0.45\textwidth}
    \hspace*{-0.3cm}
    \centering
    \raisebox{0.15cm}{%
    \includegraphics[width=\linewidth]{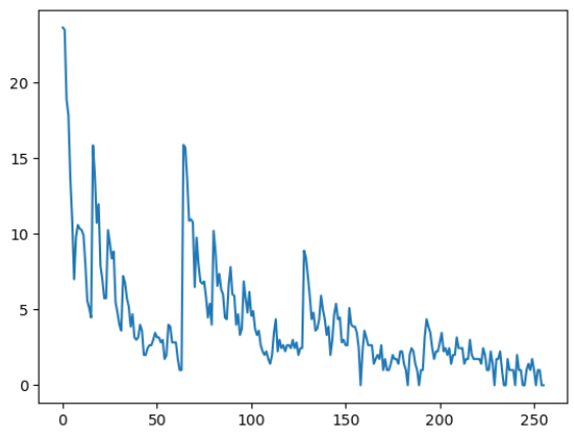}}
    \caption{\textit{ Example of state initialisation on physical backend ibm\_fez, 8000 shots}}
\end{subfigure}

\caption{\textit{State initialisation}}
\label{state init 4 figures}
\end{figure*}

\begin{figure*}
    \centering
    \includegraphics[width=0.7\linewidth]{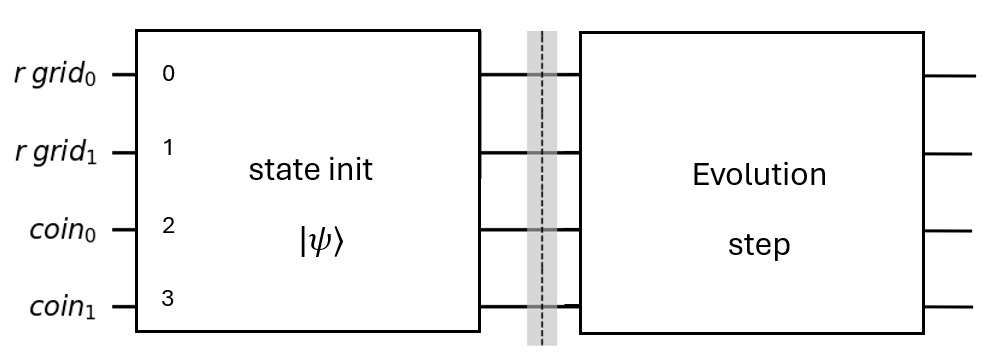}
    \caption{\textit{Initialisation of $|\psi\rangle$ at the beginning of each evolution step. Notice the order of qubits for the application of \texttt{qc.initialize()}}}
    \label{fig:placeholder}
\end{figure*}

To amplitude-encode the $1-d$ spatial discretisation of the fields in eq. (\ref{fields vector}) according to the prescription in eq. (\ref{spinor amplitude encoding}), we first establish the grid by allocating $2$ qubits to the coin register to encode 4 field components as strings of boolean values as 
\begin{equation}
|c_0 c_1\rangle \in
\begin{cases}
|00\rangle = a_r \\
|01\rangle = K_R \\
|10\rangle = nr  \\
|11\rangle = K_T
\end{cases}
\end{equation}
and $\log_2 m$ qubits to the spatial grid register to encode $m$ grid positions as 
\begin{equation}
|x_0 ... x_{m-1}\rangle \in
\begin{cases}
|0 ... 0\rangle =  r_0 \\
\vdots \\
|1 ... 1\rangle = r_{m-1}
\end{cases}
\end{equation}
so that we can encode the fields spatial discretisations in a spinor of the form
\begin{equation}
|\Psi\rangle=\psi_{c_0c_1}(r)|c_0 c_1\rangle|x_0...x_{m-1}\rangle 
\label{psi}
\end{equation}
where the boolean value of $\left|c_0 c_1\right\rangle$ select the field component, that of $\left|x_0 \ldots x_{m-1}\right\rangle$ selects a given position on the spatial grid, and the amplitude $\psi_{c_0 c_1}(r)$ corresponds to the value of the given field component at the specified spatial grid potition, up to a normalisation. \\

Hence, we can state-initialise the coin and grid qubits with a unit-state vector of field values discretisations
\begin{equation}
\begin{aligned}
\psi_{c_0 c_1}(r)= A[\, &a_r(x_0), \ldots, a_r(x_{m-1}),
K_R(x_0), \ldots, K_R(x_{m-1}), \\
&n_r(x_0), \ldots, n_r(x_{m-1}),
K_T(x_0), \ldots, K_T(x_{m-1})\, ]
\label{encoding order}
\end{aligned}
\end{equation}
where $a_r\left(x_0\right)$ is the value of $a_r$ at the grid position $x_0$ etc., and A is a global normalisation factor such that the sum of the amplitudes squared is equal to 1. 

Notice that this basis is rotated with respect to the ordering of fields in \ref{fields vector}. This choice makes the operations on the coin in the advection step in the following section (\ref{advection step}) particularly straigntforward. However, it is possible to use the original basis and perform permutations the components (when necessary at any point in the quantum circuit) by applying a Pauli-$X$ gate on either of the coin qubits (depending on the desired change in the boolean variables), since 
\begin{equation}
X|0\rangle=\left(\begin{array}{ll}
0 & 1 \\
1 & 0
\end{array}\right)\binom{1}{0}=\binom{0}{1}=|1\rangle
\end{equation}
and 
\begin{equation}
X|1\rangle=\left(\begin{array}{ll}
0 & 1 \\
1 & 0
\end{array}\right)\binom{0}{1}=\binom{1}{0}=|0\rangle
\end{equation}

It is important to notice that the discretised field values are encoded in amplitudes, but upon measurement of the qubits we obtain amplitudes squared, so any sign information, while being taken into account by the algorithm components (statevector is initialised with signed data at each step), is lost upon measurement and needs to be retrieved either with an additional sign register (cf. data loading methods in refs. \cite{Harrow_2009, dervovic2018quantumlinearsystemsalgorithms, saito2021iterativeimprovementmethodhhl, Pellow-Jarman_Sinayskiy_Pillay_Petruccione_2023b, ali2026solvingsystemslinearequations}), or inferred by considering critical points of the data.\\

Finally, it is important to warn the reader that Qiskit employs the little-endian bit ordering in state preparation and measurement, which may seem counterintuitive.

The means that while the labelling of qubits in the quantum circuit increases from top to bottom, the qubits importance (and hence the tensor structure) increases from bottom to top (cf. fig. \ref{state init 4 figures}). This means that the strings of boolean values obtained from measuring a circuit should be read right to left, with the rightmost bit corresponding to $c_0$ in our formulation (\ref{psi}) but to the qubit labelled as \textit{coin}$_1$ from the circuit. Because of this, notice the numbers within the operators applied in the circuit components, which indicate to what qubit order to apply them, increase from bottom to top, while the state preparation and measurement use the reverse ordering. 
Finally, it is important to notice that the \textit{state-initialisation} or \textit{state-preparation} subroutine is a unitary quantum sub-circuit that can be obtained in Qiskit from the \texttt{qc.initialize()} command, where 
\texttt{qc=QuantumCircuit(n)} and $n$ is the number of qubits.

\subsection{Advection step}
\label{advection step}

In this section, we focus on the quantum computing implementation of the following part of eq. (\ref{WEBB evolution}):
\begin{equation}
\boldsymbol{C}^{\hat{r}} B_R \partial_r \mathbf{q}
\label{advection}
\end{equation}
$B_R$ sets the initial relationship of coordinate radius to proper radius, and for simplicity and without loss of generality, we set $B_R =1$ (alternatively it can be implemented using the same subroutine as the static lapse function that we will describe in the following section \ref{fixed lapse section}), and work in a gauge where $A_r = 0 $ (see Appendix \ref{appendix implementing} section on implementing $A_r \not= 0 $), so that 
\begin{equation}
\boldsymbol{C}^{\hat{r}}=-\left(\begin{array}{cccc}
0 & 1 & 0 & 0 \\
1 & 0 & 0 & 0 \\
0 & 0 & 0 & 1 \\
0 & 0 & 1 & 0
\end{array}\right)
\label{cr}
\end{equation}
which can be diagonalised as $
P^{-1} \boldsymbol{C}^{\hat{r}} P=\operatorname{diag}(-1,1,-1,1)$ with 
\begin{equation}
P=\left(\begin{array}{cccc}
1 & 1 & 0 & 0 \\
1 & -1 & 0 & 0 \\
0 & 0 & 1 & 1 \\
0 & 0 & 1 & -1
\end{array}\right)
\label{P}
\end{equation}
which corresponds to two eigenmodes with amplitude $a_{\hat{r}}+K_R$ and $n_{\hat{r}}+K_T$ and eighenvalue $-1$, and two with amplitude $a_{\hat{r}}K_R$ and $n_{\hat{r}}-K_T$ and amplitude $+1$.
Notice that in eq. (\ref{evol schw}), $\boldsymbol{C}^{\hat{r}} B_R \partial_r \mathbf{q}$ is on the left hand side, while to obtain the update rule for a single timestep, we will need to move it to the right hand side, which changes the sign of the eigenvalues.\\

\begin{figure}
    \centering
    \includegraphics[width=\linewidth]{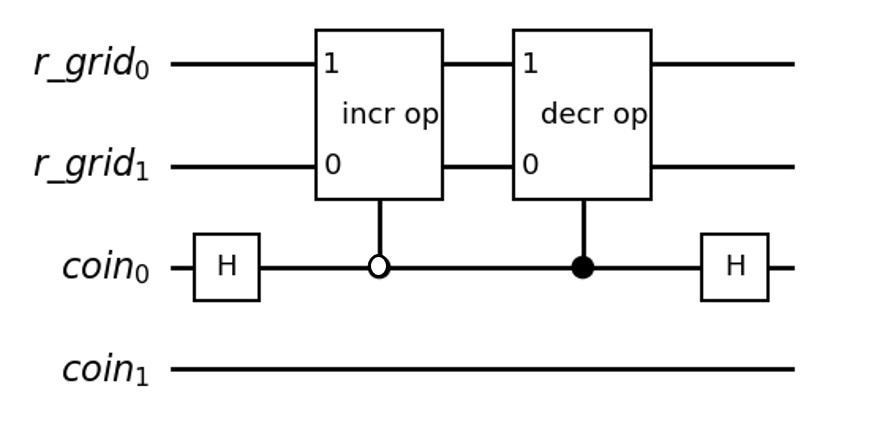}
    \caption{\textit{ Advection step. The white dot corresponds to conditioning the operation on $\text{coin}_0 = |0\rangle$, while the black dot corresponds to conditioning on $\text{coin}_0 = |1\rangle$. Notice in practice only the black dot control is native in Qiskit, so to obtain a white-dot control, apply an $X$ gate before and after a black-dot control to obtain a white-dot control.}}
    \label{advection step figure}
\end{figure}

\begin{figure}
    \centering
    \includegraphics[width=\linewidth]{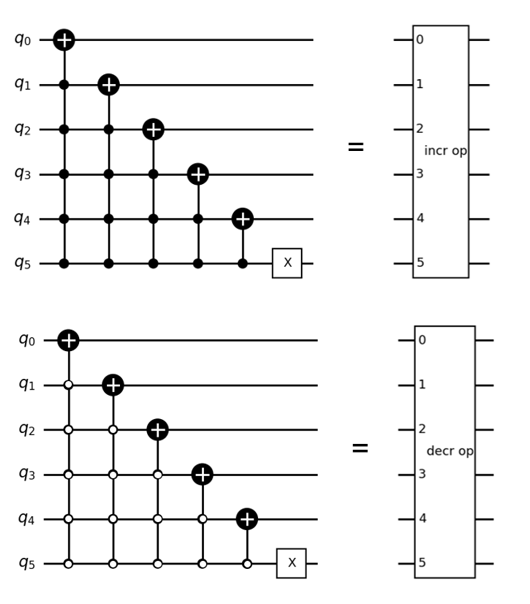}
    \caption{\it The implementation of the derivative operator used in our algorithm  $\partial_r$ is borrowed from ref.\cite{Fillion-Gourdeau_MacLean_Laflamme_2017} and \cite{Douglas_Wang_2009}.}
    \label{incr decr}
\end{figure}

In terms of quantum gates, given the ordering of fields in eq. (\ref{encoding order}), we can obtain the required eigenvectors simply by applying 
\begin{equation}
P=P^{-1} = \sqrt{2}(I \otimes H)
\end{equation}
to the coin register, where $H$ is the Hadamard gate
\begin{equation}
H=\frac{1}{\sqrt{2}}\left(\begin{array}{cc}
1 & 1 \\
1 & -1
\end{array}\right)
\end{equation}
After we have moved to the eigenvectors basis (cf. fig.\ref{advection step figure}), we apply the shift operators (``increase" and ``decrease" operators (fig. \ref{incr decr}) that shift each field value discretised on the grid to the neighbouring grid position to the right and left respectively) controlled controlled on the coin register, and then we move back to the original basis by applying again $P$.

To code the controlled shift operator in Qiskit, we first build the left and shift (or \textit{increase and decrease}) operators as subcircuits as shown in fig.\ref{incr decr}. Then, the subcircuits can be turned into custom gates through the 
\texttt{qc.to\_gate()} command, and made controlled through the \texttt{.control(n)} command, where $n$ is the number of control qubits.

Finally, notice that a discrete quantum walk with $\Delta r = \Delta t$ saturates the CFL bounds for the underlying advection system. Furthermore, notice that amplitude blow up is prevented from the Hadamard being unitary and the shift operators being permutations. The full advection step is unitary and hence stable by construction. Finally, notice that for more general systems, the dispersion relations for the propagation of modes can be tuned by substituting the $H$ gate with a more general rotation matrix as explained in section \ref{CFL}, and the shift operators can be adapted to alternative lattice structures of the grid.

\begin{algorithm}[H]
    \caption{Advection step}
    \label{alg:advection}
    \begin{algorithmic}[1]
        \State Rotate the coin register to eigenvectors basis where $C^r$ is $\operatorname{diag}(1,-1,1,-1)$;
        \State Perform advection through increase/decrease gates controlled on coin register;
        \State Rotate coin register back.
    \end{algorithmic}
\end{algorithm}

\subsection{Collision step}

In this section, we focus on modeling source vector $S$, whose components are non-linear functions of the four fields
\begin{equation}
\small
\begin{aligned}
\mathbf{S}
&=
\mathbf{f}
\left(
\begin{bmatrix}
K_R \\
a_r \\
K_T \\
n_r
\end{bmatrix}
\right)=
\begin{bmatrix}
f_1(K_R,a_r,K_T,n_r) \\
f_2(K_R,a_r,K_T,n_r) \\
f_3(K_R,a_r,K_T,n_r) \\
f_4(K_R,a_r,K_T,n_r)
\end{bmatrix}=
\\[0.5em]
&=
\begin{bmatrix}
a_r^2 - n_r^2 - K_R^2 + K_T^2 + \dfrac{2n_r}{R}
\\[0.4em]
\dfrac{2(K_R-K_T)}{R}
\\[0.4em]
\dfrac{a_r+n_r}{R}
+ K_RK_T - K_T^2 - n_r^2 - a_{\hat{\mu}}n_r
\\[0.4em]
\dfrac{K_T-K_R}{R}
- a_rK_T + n_r(K_R-2K_T)
\end{bmatrix}
\end{aligned}
\end{equation}

\noindent $S$ is the non-linear part of the equation and it can be implemented in different ways, depending on the regime of the physics problem we are looking at, level of accuracy and efficiency required. The first step consists on finding an operator, $M_S$, implementing exactly or approximately $S(q)$. Three methods we considered, of decreasing exactness and increasing efficiency, are: 
\begin{enumerate}
    \item Carleman linearisation;
    \item local linearisation through the Jacobian matrix of $S(q)$;
    \item heuristic collision operator; 
\end{enumerate}
We give more details of each method in the Appendix section \ref{appendix collision}. For this proof-of-principle implementation of the algorithm, we tested the Jacobian and heuristic collision operators and checked that they provide similar results (notice figure \ref{QNMs convergence test} shows the result obtained from the heuristic collision operator in the Appendix, while fig.\ref{jacobian vs heuristic} compares signal extraction from the Jacobian and heuristic collision operators).

It is worth briefly commenting on why those methods work while not being exact like the Carleman linearization method, since such methods might produce matrices with a sparse or local structure that can approximate the non-linear dynamics  more efficiently than the more general Carleman linearisation method. 

Let a PDE contain a nonlinear source term 
\begin{equation}
\frac{\partial u}{\partial t}=S(u) .
\end{equation}
with Jacobian
\begin{equation}
J(u)=\frac{\partial S}{\partial u}
\end{equation}
A first-order Taylor expansion around a reference state $u_*$ gives
\begin{equation}
S(u) \approx S\left(u_*\right)+J\left(u_*\right)\left(u-u_*\right) .
\end{equation}
Notice that if $u_*$ is fixed, then the source term and the whole evolution has indeed been linearized. However, in schemes where the Jacobian is continuously updated ($J=J\left(u^n\right)$ where $u^n$ is the current solution), while each individual step uses a linear operator, the operator itself depends on the previous step solution, so the overall method remains nonlinear since $J\left(u^n\right) \neq \text{const}$ constant. This is analogous to Newton's method and implicit PDE solvers. In particular, this applies to out implementatio of the coin operator from the Jacobian and heuristic methods, since we use the Jacobian matrix entries as couplings among the fields in our evolution (cf. Appendix section \ref{appendix collision}). It is important to stress that we do not use the Jacobian method directly, which would require one to compute the Jacobian classically and implement it as a quantum operator at each step. Instead, we use the Jacobian to derive the couplings between the fields that enter the effective collision operator. We build the heuristic collision operator in a similar way as a sparsified version of it. This means that the collision operator can be built at the beginning of the simulation and does not need to be updated, and hence is fully quantum mechanical (differently from alternative methods to implement nonlinear dynamics in quantum computing, such as VQE, in hybrid classical-quantum algorithms\cite{McArdle_2019, wang2026reviewvariationalquantumalgorithms}). \\

\begin{figure}[H]
     \centering
     \includegraphics[width=1\linewidth]{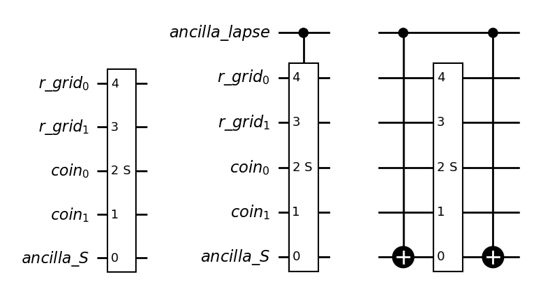}
     \caption{\it Application of the $U_S(t)$ operator and two equivalent ways to control it on the lapse (cf. section \ref{fixed lapse section})}
     \label{collision and its controls}
 \end{figure}

While each of the methods mentioned above would give a different $M_S$, the common feature is that if $M_S$ is not unitary, we embed it in the larger block-diagonal matrix 
\begin{equation}
H_S = \begin{pmatrix}
0 & M_S \\
M_S^\dagger & 0
\end{pmatrix}
\end{equation}
which is Hermitian and can be implemented as a Hamiltonian gate $U_S(t)=e^{-i t H_S}$ applied to the an ancilla qubit (notice augmented Hilbert space needed due to dimension of $H_S$), the coin and grid registers (cf. fig. \ref{collision and its controls}).

To code the collision operator in Qiskit, we use \texttt{HamiltonianGate(Operator(H), time = t)}, where $H=H_S$ and $t=\Delta t$ for a single timestep. When running the algorithm on physical quantum computers, it is convenient to first decompose $H_s$ into Pauli gates using \texttt{SparsePauliOp.from\_operator(Operator(H))} which decomposes the matrix into a weighted sum of Pauli operators $H=\sum_k c_k P_k$, where $P_k$ are strings of Pauli operators (like "IXY...Z" of length tr$H_S$) and $c_k$ are complex coefficients. We then use this decomposition to implement the Trotter steps $e^{-i H t} \approx \prod_k e^{-i \theta_k P_k t}$, where $\theta_k=c_k\cdot \frac{\Delta t}{\text {num\_slices }}$.

\begin{algorithm}[H]
\caption{S term}
\label{alg:S}
\begin{algorithmic}[1]
\Require 1 ancilla qubit
\State Compute Jacobian matrix;
\State Build scattering matrix from Jacobian matrix;
\State Use Hermitian operator trick to encode scattering matrix into Hermitian operator $H_S$;
\State Apply $e^{-iH_St}$ to ancilla, coin and grid qubits.
\end{algorithmic}
\end{algorithm}

\subsection{Static lapse and shift subroutines}
\label{fixed lapse section}

We implement a lapse function $\alpha(r)$ that remains static during the timestep as follows. We first amplitude-encode the lapse function $\alpha(r) \equiv \alpha_r$ onto an ancilla qubit $q_a$; then conditioning a given operation on $q_a = |1\rangle_a$ corresponds to pre-factoring such operation by a function of $\alpha_r$. Finally, we undo the amplitude-encoding of the lapse function on the ancilla to disentangle the ancilla qubit $q_a$ from the grid qubits. We go through the details of each step below.\\

\begin{figure}
    \centering
    \includegraphics[width=1\linewidth]{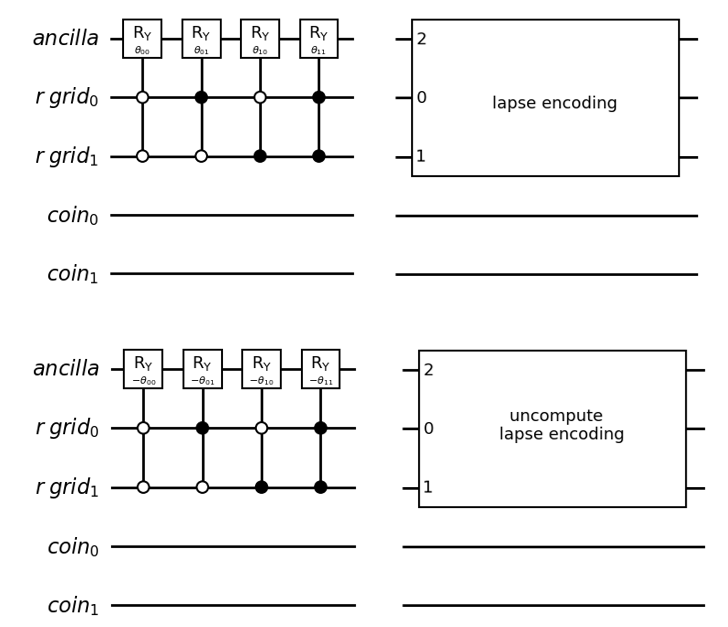}
    \caption{\it Lapse encoding and disentangling subroutines. cf. ref. \cite{Budinski_2021} which uses a similar implementation.}
    \label{lapse encoding figure}
\end{figure}

\begin{figure*}[t]
\centering

\includegraphics[width=0.9\linewidth]{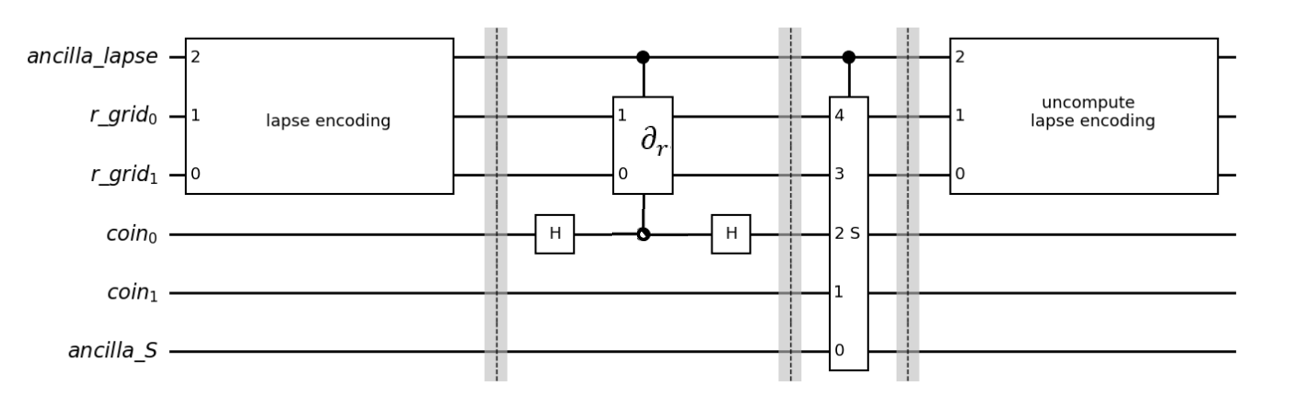}
\caption{\it Advection and collision operations controlled on lapse ancilla qubit. Notice collision operation can alternatively be controlled as shown in fig.\ref{collision and its controls}.}
\vspace{0.5cm}
\label{fixed lapse subroutine}

\includegraphics[width=0.75\linewidth]{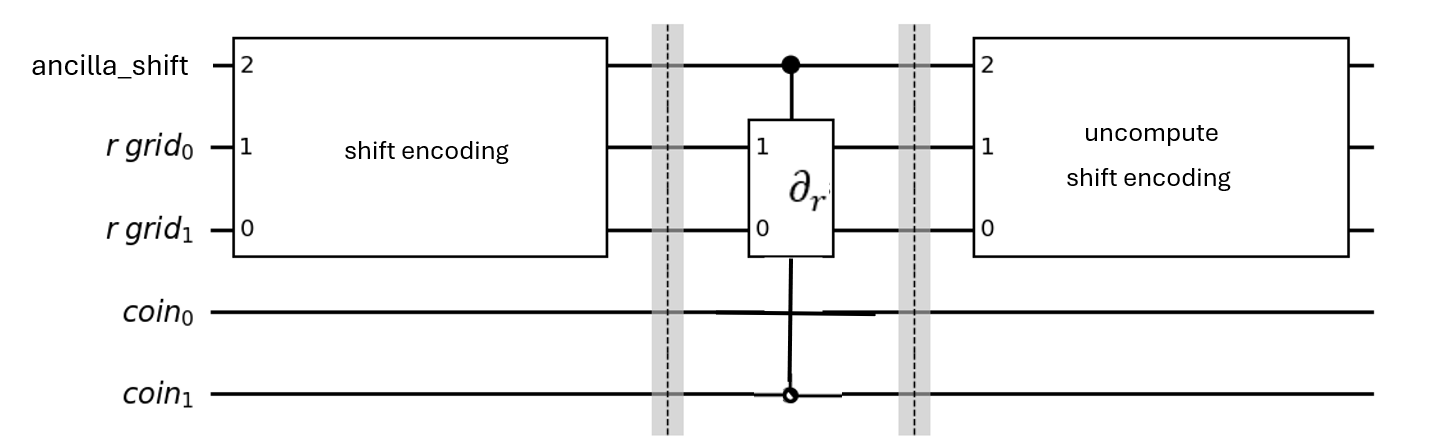}

\caption{\it Shift subroutine}
\label{shift fig 8}
\vspace{0.5cm}

\includegraphics[width=\linewidth]{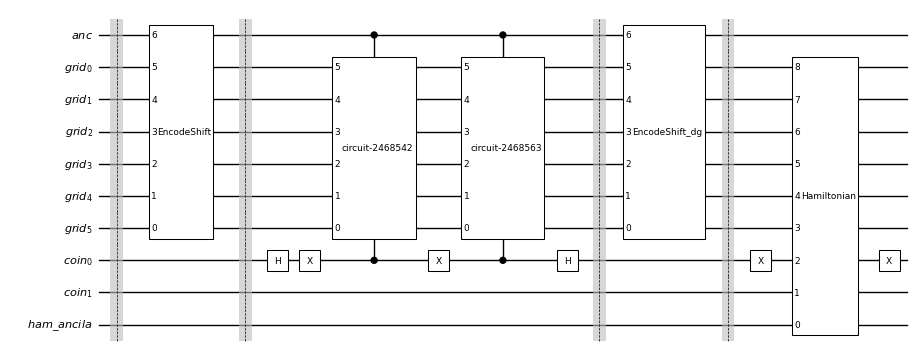}

\caption{\it Circuit used for the Qiskit simulation.}
\label{codeed circuit}

\end{figure*}

\subsubsection{Pointwise lapse encoding}

This section is about implementing the lapse and shift functions in 
\begin{equation}
D_0=\frac{1}{\alpha}\left(\partial_t-\beta^r \partial_r\right)
\end{equation}
A fixed lapse function $\alpha_r$ can be amplitude-encoded in an ancilla qubit through controlled $R_y(\theta_r)$ rotations conditioned on the spatial grid register. The amplitude-encoding relation is given by 
\begin{equation}
U_{encode} = R_y\left(\theta_r\right)|0\rangle_a|r\rangle=(\sqrt{1-p_r}|0\rangle_a+\sqrt{p_r}|1\rangle_a)|r\rangle
\label{lapse encoding}
\end{equation}
where $|r\rangle = \left|x_0 \ldots x_{m-1}\right\rangle_r$, $p_r=\alpha_r$ defines a set of encoding weights over the grid. The angles $\theta_r$ are found according to the following prescription
\begin{equation}
\sin \left(\frac{\theta_r}{2}\right)=\sqrt{p_r} 
\end{equation}
Notice that for each $r \in \{0,..,m-1\}$, the rotation $R_y\left(\theta_r\right)$ is controlled on the $m$ grid qubits being in the boolean combination corresponding to $r$.

Finally, notice that this construction directly works for our choice of lapse function for this toy model demonstration (fig. \ref{chosen lapse}), and for any bounded $\alpha_r \in[0,1]$, since it is fully expressible via the sine mapping in $R_y\left(\theta_r\right)$. This construction can be extended to a generalised lapse function by using\begin{equation}
p_r=\frac{\alpha_r^2}{\sum_{r_0}^{r_{m-1}} \alpha_r^2}
\end{equation} in which case $p_r$ can be interpreted as a probability distribution.

\subsubsection{Condition operations on lapse ancilla qubit}

Conditioning operations on $q_a=|1\rangle_a$ now induces an $\alpha_r$-dependent weighting in the reduced dynamics after tracing out or postselecting the ancilla. Disentangling the ancilla from the grid register as explained in the next section reduces entanglement-induced mixing effects when tracing out the ancilla and is hence advised.
The advection step is controlled on $q_a=|1\rangle_a$ as shown in fig.\ref{fixed lapse subroutine}. The collision step can be controlled on $q_a=|1\rangle_a$ in multiple ways: i) controlling the whole operator on the ancilla (which is however costly); ii) controlling ancilla$_S$ on ancilla$_a$ (cf. fig.\ref{collision and its controls}). Finally, notice that an alternative option is to apply the uncontrolled collision operator before or after uncomputing the lapse. This leads to a non-commuting operator ordering, where leading-order deviations are governed by the corresponding commutator and the dynamics of $S$ is only effectively weighted by $\alpha_r$.

\subsubsection{Uncomputing the lapse encoding}

Notice that at this point $q_a$ is still entangled with the grid qubits. To disentangle it, we apply $U_{encode}^{\dagger}$ which amounts to applying the same subroutine as before with negative angles (cf. fig. \ref{lapse encoding figure}).

To code the lapse encoding step in Qiskit, we build the subcircuit shown in fig.\ref{lapse encoding figure} and then wrap it into an Instruction object through the 
\texttt{qc.to\_instruction()} command to easily reuse it in the main circuit (cf. fig.\ref{codeed circuit}). To obtain the inverse of the lapse encoding we simply apply \texttt{encode\_gate.inverse()}.

\begin{algorithm}[H]
\caption{Static lapse}
\label{alg:fixed_lapse}
\begin{algorithmic}[1]
\Require 1 ancilla qubit
\Require discretised lapse function $\alpha$
\State Obtain $\theta_{x_m}$ values from $\alpha$
\For {$i = 0$ to $m$}
\Statex rotate ancilla using $RY(\theta_{x_i})$ controlled on grid qubits boolean value corresponding to grid position $x_i$;
\EndFor    
\State Control required operations on ancilla qubit;
\For {$i = 0$ to $m$}
\Statex rotate ancilla using $RY(-\theta_{x_i})$ controlled on grid qubits boolean value corresponding to grid position $x_i$.
\EndFor
\end{algorithmic}
\end{algorithm}

\subsubsection{Static shift}

The subroutine to implement a shift $\beta^r \partial_r$ with a static shift function $\beta^r$ is the same as for the lapse, but the derivative $\partial_r$ is implemented with either the increase or decrease operator (fig.\ref{incr decr}) depending on the direction, and controlled as shown in fig.(\ref{shift fig 8}), i.e. conditioned on coin$_1 =|0\rangle$ and coin$_1 =|1\rangle$ as it needs to be to all fields equally.

\begin{algorithm}[H]
\caption{Static shift}
\label{alg:fixed_shift}
\begin{algorithmic}[1]
\Require 1 ancilla qubit
\Require discretised shift vector $\beta^r$
\State Obtain $\theta_{x_m}$ values from $\beta^r$
\For {$i = 0$ to $m$}
\Statex rotate ancilla using $RY(\theta_{x_i})$ controlled on grid qubits boolean value corresponding to grid position $x_i$;
\EndFor 
\State Control required operations on  ancilla qubit;
\For {$i = 0$ to $m$}
\Statex rotate ancilla using $RY(-\theta_{x_i})$ controlled on grid qubits boolean value corrresponding to grid position $x_i$.
\EndFor 
\end{algorithmic}
\end{algorithm}

Finally, in the Appendix section \ref{appendix dynamical lapse} we comment on alternative methods to implement dynamical lapse and shift functions.

\subsection{Boundaries}

For the problem at hand, we need to define an inner boundary located at the black hole horizon ($R\rightarrow 1$) and an outer boundary ($R\rightarrow \infty$ but effectively $R\rightarrow R_{max}$).\\

In the Schwarzschild black hole example in \cite{Buchman_2005} which we use to test our proof-of-principle quantum algorithm, the inner boundary is treated as an \textit{excision boundary} (an artificial boundary located inside of the black hole horizon, the region inside of which it is simply removed from the computational domain). Since all physical and characteristic modes already propagate inward there (toward the singularity), no physical information can emerge from the inner boundary. In other words, the Cauchy nature of the problem together with the causal structure make it so that it is not necessary to further impose a dynamical inner boundary and the information needed to updated the field values there can be obtained from boundary values or using upwind differencing. At the outer boundary, outgoing characteristic modes are evolved freely, while the incoming modes are either fixed analytically to the Schwarzschild solution, or minimized to avoid spurious reflections. \\

In practice, since the default grid in Qiskit is periodic, we need to make sure that there is not spurious reflection or transmission happening at grid points $x_0$ and $x_{m-1}$. In practice, it is sufficient to impose $x_{m-1} = x_{m-2}$ before re-initialising the data at each step to obtain an absorbing boundary at $R_{max}$ and leave the dynamical evolution free at $x_0$.

Alternatively, one can create \textit{sponge} boundaries by setting the grid in such a way as to allow for extra points to the left of $x_{min}$ and to the right of $x_{max}$, and initialising the field values in those segments in such a way as to implement the required boundary conditions. As an example, it may be possible to initialise the field values in the extra grid segment towards $R_{max}$ as being zero, and the field values in the extra grid segment before the horizon with boundary values.\\

Finally, notice that if we only measure the system grid, the amplitudes will sum to $1$ regardless to whether the amplitude has leaked outside of the system. So, to obtain a meaningful result, we have to also either postprocess the result after each step by subtracting the amplitude squared of the fields at the external boundary (or sponge boundary segments) from the norm of the fields, or implement a quantum \textit{amplitude leakage} operator (cf. Appendix section \ref{appendix boundary}).

\begin{figure}
    \centering
    \includegraphics[width=0.9\linewidth]{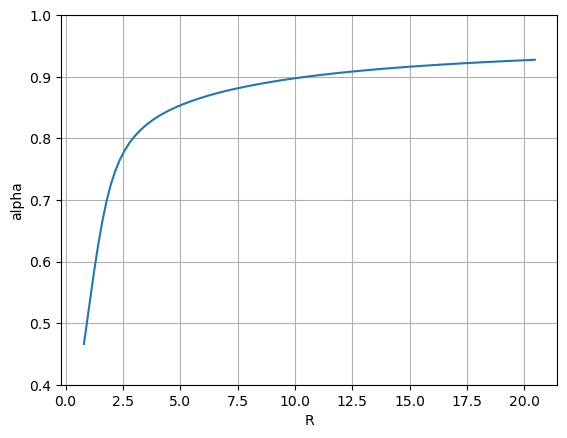}
    \caption{\it Lapse function chosen for testing the algorithm}
    \label{chosen lapse}
\end{figure}

To test the algorithm, we evolve $K_R, a_r, K_t$ and $n_r$ through the quantum circuit in fig. \ref{codeed circuit}, where $S$ is implemented through the collision operator in Appendix section \ref{appendix collision}, where we also show how the heuristic and Jacobian methods work similarly, and $S$ can safely be positioned after the uncomputing of the lapse encoding. We choose $\beta=0$ and lapse $\alpha$ given by fig.\ref{chosen lapse}

We simulate the quantum circuit classically using the statevector option of the Qiskit aer-simulator (in what follows we refer to it as \textit{Qiskit noiseless or ideal simulation to distinguish it from runs on physical backends}), and use the evolved values to compute $\Psi_4$ (eq. \ref{Psi 4 qnms approx}). 

\begin{figure*}[t]
    \hspace{25mm}
    \includegraphics[width=0.8\linewidth]{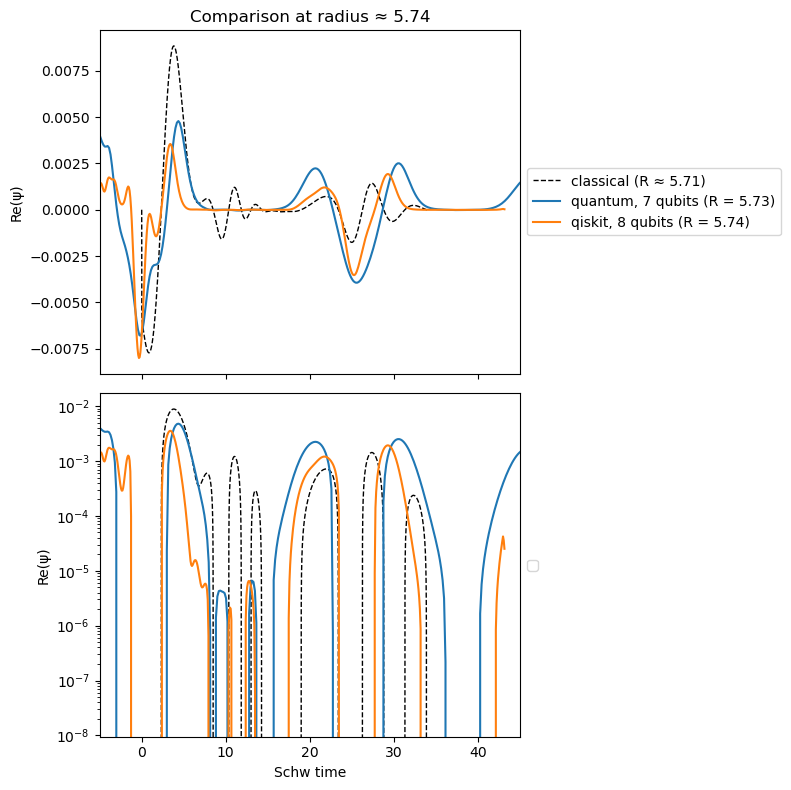}
    \caption{\it Comparison of Re$(\Psi_4)$ signals extracted at the same radius $R$ from two ideal noiseless Qiskit simulations ($m = 7,8$ grid qubits, $\Delta t = \Delta R = 0.14  \text{ and } 0.07$ respectively in $M=1/2$ geometric units), and from classical finite difference method ($\Delta t = 0.0025, \Delta R = 0.14$). The convergence testing we have been able to perform so far shows the signal peaks moving towards the classical solution as we increase the number of qubits and hence the resolution.}
    \label{QNMs convergence test}
\end{figure*}

Fig.\ref{QNMs convergence test} shows the obtained results for $m = 6$ 6 and 7 grid qubits ( $\Delta t = \Delta R = 20/2^{m}$ in $M=1/2$ geometric units), compared against the result obtained from classical finite difference method.

For the Qiskit and classical runs, it is possible to limit the fit to the ringdown segment of the signal. However, this is not possible for the runs on physical quantum backends due to noise. Because of this, for the latter we apply a bootstrap fitting method, where we fit the whole signal over multiple bins. For the classical and Qiskit runs, we apply the same method but for each radial bin we track where the ringdown segment starts and only fit after that time. The results are shown in table \ref{results table}.

\begin{figure*}[t]
\centering
\begin{subfigure}{1\columnwidth}
    \centering
    \includegraphics[width=\linewidth]{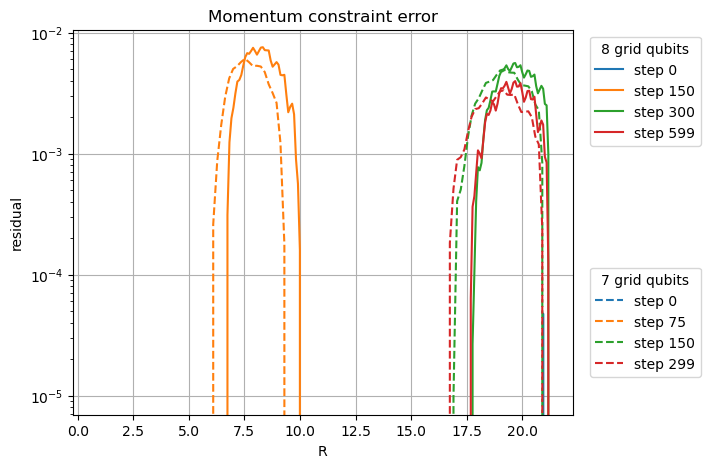}
    \caption{\it Momentum constraint (eq. \ref{momentum constraint}) error for noiseless ideal Qiskit simulations of 7 and 8 qubits.}
\end{subfigure}
\hfill
\begin{subfigure}{1\columnwidth}
    \centering
    \includegraphics[width=\linewidth]{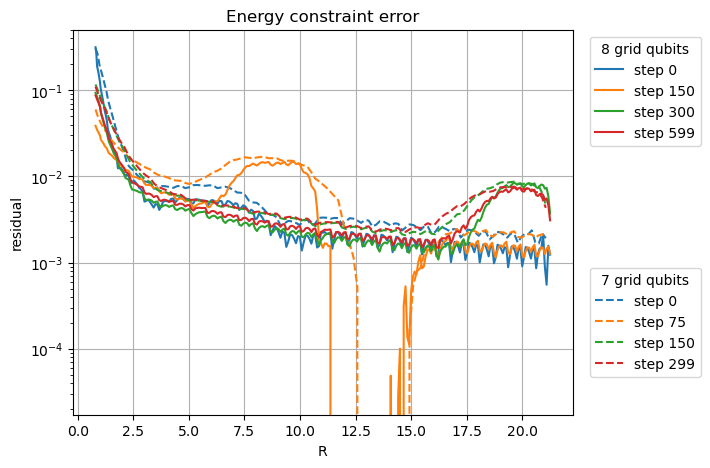}
    \caption{\it Energy constraint (eq.\ref{energy constraint}) error for noiseless ideal Qiskit simulations of 7 and 8 qubits.}
\end{subfigure}
\caption{\it Momentum and energy constraints (eqs. \ref{momentum constraint}-\ref{energy constraint}) error for noiseless ideal Qiskit simulations of 7 and 8 qubits. Notice imprint of initial perturbation remains controlled. Initial energy constraint violation may be responsible for system relaxing towards solution at the very start of simulation. Also, notice late time reflection at outer boundary.}
\end{figure*}

\begin{figure*}[t]
\centering
\begin{subfigure}{1\columnwidth}
    \centering
    \includegraphics[width=\linewidth]{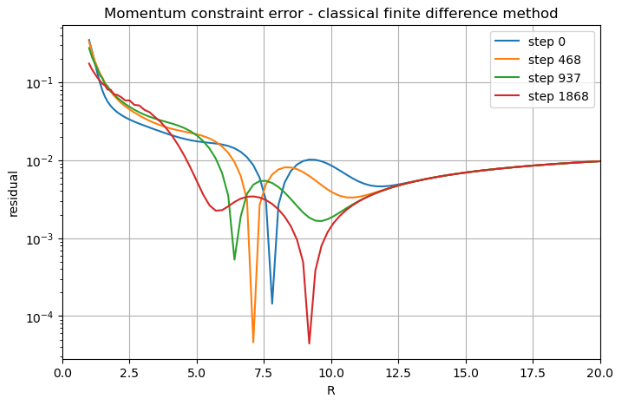}
    \caption{\it Momentum constraint (eq. \ref{momentum constraint}) error for  for classical finite difference method}
\end{subfigure}
\hfill
\begin{subfigure}{1\columnwidth}
    \centering
    \includegraphics[width=\linewidth]{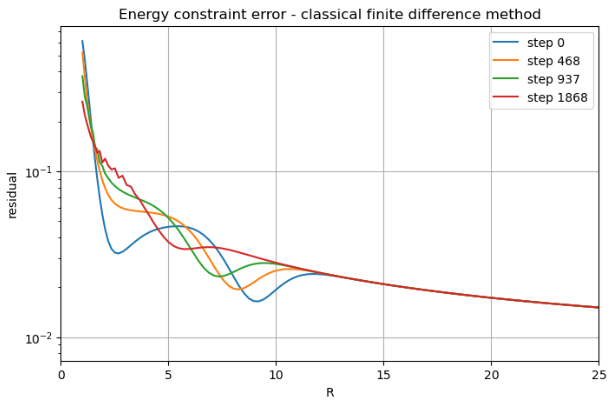}
    \caption{\it Energy constraint (eq.\ref{energy constraint}) error for classical finite difference method}
\end{subfigure}
\caption{\it Momentum and energy constraints (eqs. \ref{momentum constraint}-\ref{energy constraint}) error for classical finite difference method. Notice the errors are comparable in size for the classical and quantum simulations, and the peaks are located approximately at the same values of $R$ (the number of steps has been chosen taking into account the different $\Delta t$) to compare at the same simultation time. Notice the errors from the classical simulation lack the peaks corresponding to spurious reflection due to a larger radial grid ($R_{max}\approx 80$).}
\end{figure*}

\section{Running the algorithm on physical backends}
\label{running algorithm}

\begin{figure*}[t]
    \hspace{3cm}
    \includegraphics[width=0.8\linewidth]{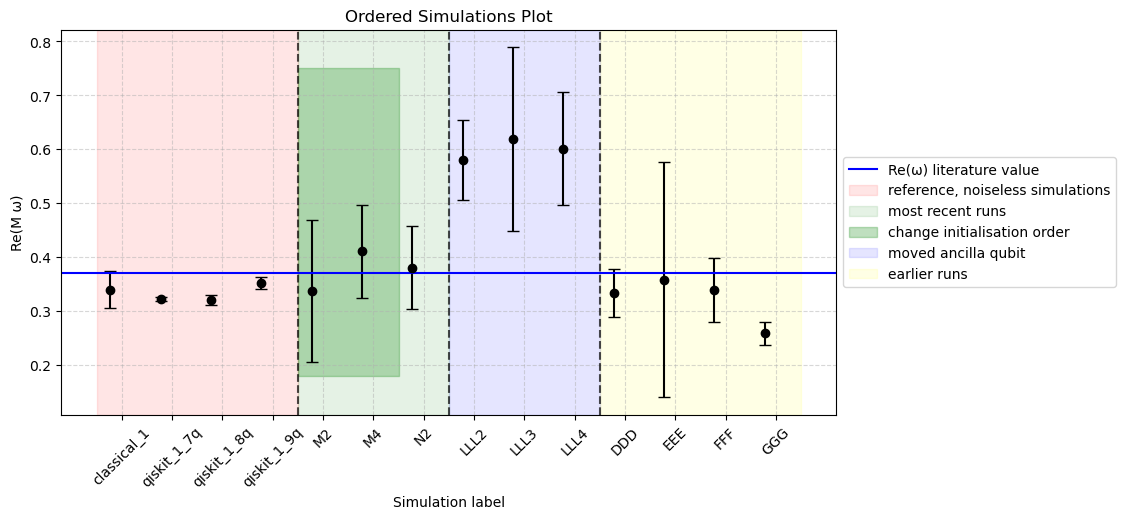}
    \caption{\it QNMs frequencies extracted from various runs on clssical simulator through Qiskit and physical quantum backends.}
    \label{frequencies extracted plot}
\end{figure*}

\begin{table*}[t]
\begin{tabular}{|l|l|c|c|c|c|c|c|c|}
\hline
run name & backend & Shots & opt. level  & grid qubits & init order & ancilla order & circuit & notes \\
\hline
classical\ & na & n.a. & n.a. & n.a. & 1 & n.a. & n.a. & n.a.\\
qiskit$_{7q}$ & aer\_simulator  & n.a. & n.a. & 7 & 1 & 1 & full & n.a.\\
qiskit$_{8q}$ & aer\_simulator  & n.a. & n.a. & 8 & 1 & 1 & full & n.a.\\
R(M2) & ibmq\_fez & 8000 & 3 & 6 & 2 & 1 & full &n.a. \\
R(M4) & ibmq\_marrakesh & 8000 & 3 & 6 & 2 &1 & full &n.a.\\
R(N2) & ibmq\_marrakesh & 8000 & 3 & 5 & 1 &1 & full & n.a.  \\
A(LLL2) & ibmq\_boston & 8000 & 3 & 6 & 1 & 2 & full & n.a. \\
A(LLL3)& ibmq\_boston & 8000 & 3 & 6 & 1 & 2 &  full & n.a.\\
A(LLL4) & ibmq\_boston & 8000 & 3 & 6 & 1 & 2 &  full & n.a.\\

E(DDD) & ibmq\_kingston & 8000 & 2 & 7 & 3 & 3 & adv. only & gaussian smoothing \\
E(EEE) & ibmq\_pittsburgh & 8000 & 3 & 7 & 3 & 3& adv. only & gaussian smoothing  \\
E(FFF) & ibmq\_kingston & 8000 & 3 & 7 & 3 & 3 & adv. only &  gaussian smoothing \\
E(GGG) & ibmq\_pittsburgh & 8000 & 3 & 7 & 4 & 3& adv. only & gaussian smoothing, padding
\\

\hline
\end{tabular}
\caption{\it Details of the simulations run on physical backends from fig.\ref{frequencies extracted plot}}
\label{results table}
\end{table*}

\subsubsection{Experiment method}

Once the circuit is defined and the physical backend is selected, the circuit is transformed in a set of instructions for the backend to execute through \textit{transpiling}: this process includes converting high level gates (e.g. Toffoli, custon unitary gates, etc.) into the device's native gates, mapping logical qubits in the circuit to physical qubits in the chip, routing (when two logical qubits need a two-qubit gate but are not physically connected, SWAP operations are inserted to make it executable on the hardware topology), and optimisation (reducing circuit cost by, e.g. canceling inverse gates, merging rotations, which reduces the circuit depth). Then, operations are scheduled according to hardware timing constraints and optionally \textit{error mitigation/suppression techniques} are applied (e.g. dynamical decoupling, measurement error mitigation, zero-noise extrapolation, probabilistic error cancellation, etc.). Finally, these instructions are submitted to the physical backends for the execution, in our case, through the IBM Runtime primitive \textit{SamplerV2}, which returns the associated probability distributions after measurement and readout (where qubits states are converted to classical bits through resonator readout and signal processing).

\subsubsection{Operations decomposition: advection}

Multiply controlled gates such as those in the advection and lapse encoding steps can be synthesised in different ways with different resulting runtimes. \\

The \textit{V-chain} method uses a singly controlled V gate (such that $V^2=U$) and CNOT gates, which are recursively applied to different controls, creating a chain of intermediate controls. This method, for a gate with $k$ controls, typically uses $k-2$ ancilla qubits, circuit depth and gate count for the number of elementaty two-qubit operations grow linearly with the number of controls $k$ (i.e. $O(k)$). Such a runtime is often cheaper than ancilla-free decompositions. Fig.(\ref{advection ancilla free}) shows the runtime obtained for the ancilla-free method and Fig.(\ref{advection ancilla assisted}) compares it with the \textit{V-chain method.}\\


\subsubsection{Operations decomposition: lapse encoding}

The lapse encoding/decoding step has the structure of a \textit{uniformly controlled rotation (UCRY)}, i.e. as a sequence of controlled $R_y$ rotations arranged in a multiplexed structure.
\begin{equation}
\operatorname{UCRY}\left(\theta_0, \ldots, \theta_{2^k-1}\right)=\sum_{i=0}^{2^k-1}|i\rangle\langle i| \otimes R_y\left(\theta_i\right)
\end{equation}
This belongs to the more general category of a \textit{quantum multiplexer}, which can be synthesised using $2^k$ single-qubit $R_y$ rotations, $2^k-1$ CNOTs, ranged in a Gray-code pattern \cite{Shende_2006}. This method, for $k$ control qubits, scales as the number of possible combinations (i.e. $O(2^k)$). \\

Notice that when the lapse function $f(x)$ is structured (i.e. a defined function), it may be possible to use an alternative method: first, a \textit{reversible-arithmetic} circuit is built such that
\begin{equation}
|x\rangle|0\rangle \rightarrow|x\rangle|f(x)\rangle
\end{equation}
for which the cost is $O(\operatorname{poly}(n))$ if $N(x)$ is structured. Then, we apply a $\textit{phase kick-back}$ step such that 
\begin{equation}
|x\rangle|f(x)\rangle \rightarrow e^{-i f(x) t}|x\rangle|f(x)\rangle
\end{equation}
and finally we uncompute the first step to disentangle the ancilla 
\begin{equation}
|x\rangle|f(x)\rangle \rightarrow|x\rangle|0\rangle
\end{equation}
state-preparation. This method is efficient ($O(poly(n))$ for $n$ grid qubits) for $f(x)$ being a polynomial function, smooth functions (approximated by a low degree polynomial), or being translatable to local lattice physics (nearest neighbour interactions). We leave this for a future extension of the present work.

\subsubsection{Operations decomposition: collision operator}
A Hamiltonian acting on $n$ qubits can be \textit{Pauli-decomposed} as
\begin{equation}
H=\sum_{j=1}^M c_j P_j =\sum_j c_j\left(P_j^{(1)} \otimes P_j^{(2)} \otimes \cdots \otimes P_j^{(n)}\right) .
\end{equation}
where the coefficients $c_j$ are real numbers, $P_j$ is a tensor product of Pauli matrices (e.g., $\left.I, X, Y, Z\right)$ which forms a complete operator basis. While the computational cost for a general dense $2^n \times 2^n$ matrix can be as high as $O\left(4^n\right)$, for a sparse (or structured or local) matrix the scaling is $O(poly(n))$. \\

Since we then want to implement $U(t)=e^{-i H t}$, we use the first-order Trotter formula
\begin{equation}
e^{-i H t} \approx\left(\prod_j e^{-i c_j P_j t / r}\right)^r
\label{trotter}
\end{equation}
where $r$ is the number of Trotter steps, and eq.(\ref{trotter}) becomes exact as $r \rightarrow \infty$, and error decreases as $O\left(t^2 / r\right)$.\\

Hence, the total cost of $r$ Trotter steps is 
\begin{equation}
\text {Cost }=r \cdot M \cdot \bar{k}
\end{equation}
where $r$ is the number of Trotter steps, $M$ is the number of Pauli terms, and $\bar{k}$ is the average weight of Pauli strings. If the Hamiltonian is sparse or structured, ($M=\operatorname{poly}(n)$), it has a local interaction structure, ($\bar{k}=$ poly(1) $\quad$ or small) and a polynomial function if Totter steps is sufficient ($r=\operatorname{poly}(n, t, 1 / \epsilon)$), we have that
\begin{equation}
\text { Total cost }=\operatorname{poly}(n)
\end{equation}
\subsubsection{Measurement and potential bottlenecks}

Notice that the state-preparation for structured or continuous functions (as is the case here) can be shown to scale as $O(\operatorname{poly}(n)$ and not provide a dramatic increasing in the runtime. 
Regarding the measurement after each step, we want to measure $|\psi\rangle$ in the computational basis. Running the circuit $N$ times ("shots") lets us estimate the probabilities associated with each computational basis as 
\begin{equation}
    p_i=|\langle i \mid \psi\rangle|^2 
\end{equation}
with the statistical error scaling as 
\begin{equation}
    O\left(\frac{1}{\sqrt{N}}\right)
\end{equation}
This result could be improved in the case in which we are interested in a specific observable, e.g. a waveform, derived from the Ricci rotation coefficients evolved. It might be possible to devise a subroutine (similar to the one for the dynamical lapse in the appendix) that builds and extracts information about a given observable through \textit{soft measurement techniques} such that the evolution does not need to be stopped and the statevector containing the main evolved quantities re-initialised at each step.

\subsubsection{Experiments}

\renewcommand{\topfraction}{0.95}
\renewcommand{\bottomfraction}{0.95}
\renewcommand{\textfraction}{0.05}
\renewcommand{\floatpagefraction}{0.9}

\begin{figure*}[t]
\centering

\begin{subfigure}[t]{0.45\textwidth}
    \centering
    \includegraphics[width=\linewidth]{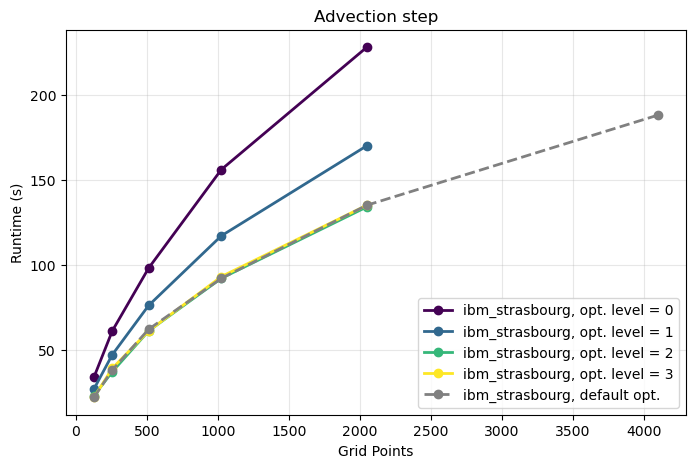}
    \caption{\it Runtime obtained from the ancilla-free decomposition of the advection step, for different optimisation options. Notice level 2 is the default optimisation option, and there is no difference between level 2 and 3 due to the circuit being already maximally optimised.}
    \label{advection ancilla free}
\end{subfigure}
\hfill
\begin{subfigure}[t]{0.45\textwidth}
    \centering
    \includegraphics[width=\linewidth]{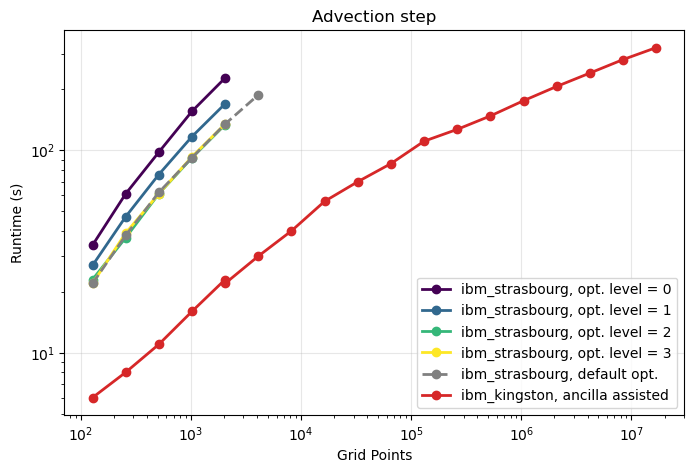}
    \caption{\it Runtime obtained from the ancilla-free decomposition compared to the V-chain decomposition, logarithmic scale.}
    \label{advection ancilla assisted}
\end{subfigure}
\caption{\it Advection runtime}
\label{advection runtime_1}
\vspace{0.3cm}

\begin{subfigure}[t]{0.45\textwidth}
    \centering
    \includegraphics[width=\linewidth]{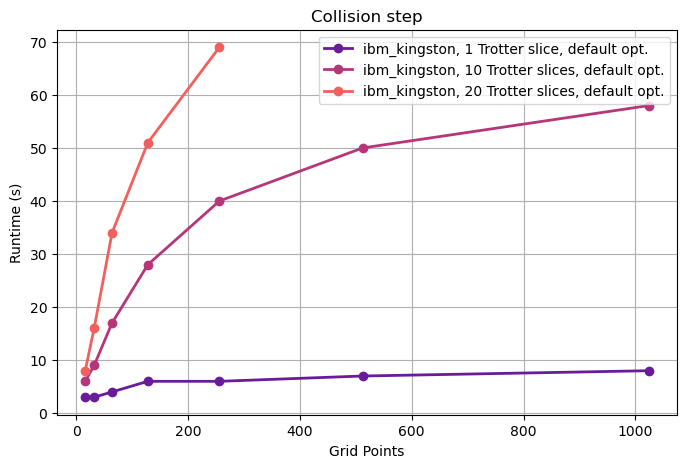}
    \caption{\it Runtime for different number of Trotter steps.}
\end{subfigure}
\hfill
\begin{subfigure}[t]{0.45\textwidth}
    \centering
    \includegraphics[width=\linewidth]{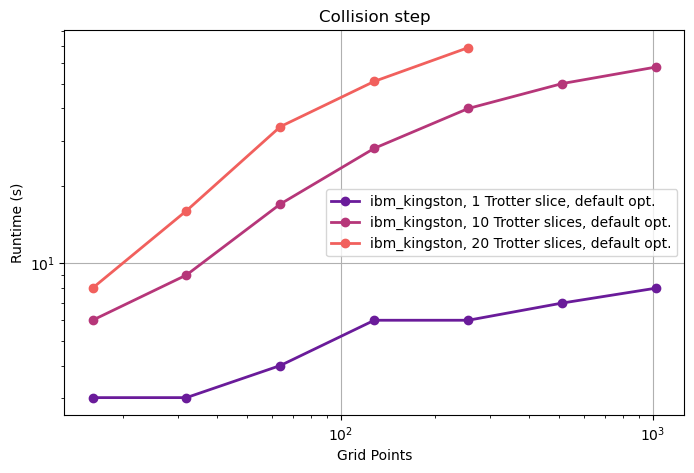}
    \caption{\it Runtime for different number of Trotter steps, logarithmic scale}
\end{subfigure}
\caption{\it Collision operator runtime}
\label{collision runtime_2}
\vspace{0.3cm}

\begin{subfigure}[t]{0.45\textwidth}
    \centering
    \includegraphics[width=\linewidth]{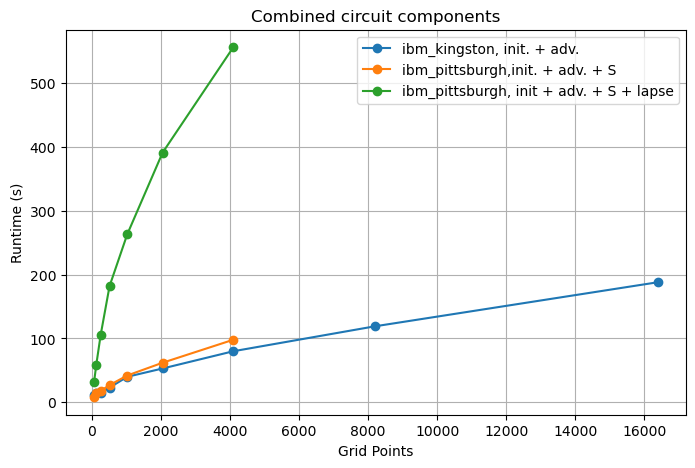}
    \caption{\it Runtime for different combination of circuit components. Notice the lapse subroutine is the most computational expensive, while initialisation, advection and S subroutines runtime is roughly logarithmic in the number of gridpoints.}
\end{subfigure}
\hfill
\begin{subfigure}[t]{0.45\textwidth}
    \centering
    \includegraphics[width=\linewidth]{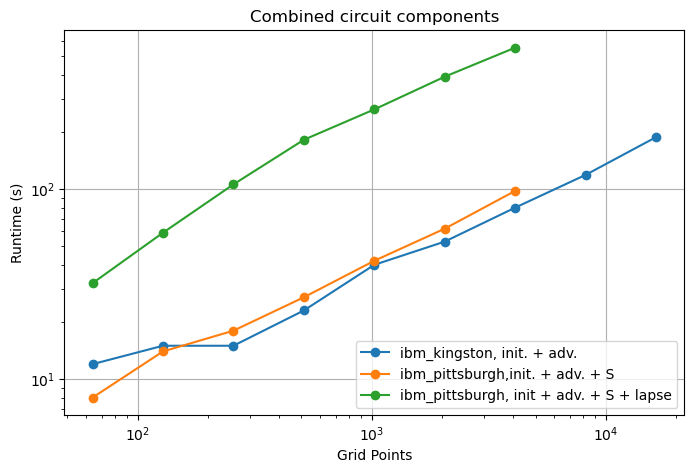}
    \caption{\it Runtime for different combination of circuit components, logarithmic scale}
\end{subfigure}
\caption{\it Runtime for different combination of circuit components}
\label{combinations runtime}
\end{figure*}

The runs on physical backends in fig.\ref{frequencies extracted plot} can be divided into three groups: 

\begin{itemize}
    \item most recent runs: where we used the circuit in fig.(\ref{codeed circuit}) without the lapse decoding step. This is because the error due to decoding would likely be larger than the error due to not disentangling the ancilla. We assign the additional ancillas due to the v-chain decomposition to qubits with the smallest $n$ following the little endian convention;
    \item runs where we changed the qubits assigned to ancillas;
    \item earlier runs, where we tested a reduced circuit (just the advection step), only evolved $K_r$ and $n_r$ and experimented with methods to reduce the noise. Also notice for these runs the initial perturbation was located towards $R_{max}$;
\end{itemize}

More details regarding qubit ordering and initialisation options are provided below and referred to in table \ref{results table}, where the specific options selected for each simulation can be found.

We test the following qubits orderings
\begin{equation}
\begin{aligned}
\text{qubits ordering 1} = {} &
[\text{ancillas}_{v\text{-}chain},
 \text{ancilla}_{lapse},
 \text{grid}_r, \\
&
 \text{coins},
 \text{ancilla}_{S}]
\end{aligned}
\end{equation}
\begin{equation}
\begin{aligned}
\text{qubits ordering 2} = {} &
[\text{ancillas}_{v\text{-}chain},
 \text{grid}_r, \\
&
 \text{coins},
 \text{ancilla}_{lapse},
 \text{ancilla}_{S}]
\end{aligned}
\end{equation}
\begin{equation}
\begin{aligned}
\text{qubits ordering 3} = {} &
[
 \text{grid}_r, 
 \text{coins},\text{ancillas}_{v\text{-}chain}]
\end{aligned}
\end{equation}

Notice moving ancilla$_{lapse}$ to qubit position $0$ means that tracing out this qubit induces more error than tracing out the ancilla located at $n_{max}$. \\

We also test the following fields initialisation orderings:
\begin{equation}
    |\psi\rangle_1 = [a_r, K_R, n_r, K_T]
\end{equation}
\begin{equation}
    |\psi\rangle_2 = [n_r, K_T, a_r, K_R]
\end{equation}
\begin{equation}
    |\psi\rangle_3 = [n_r, K_T]
\end{equation}
\begin{equation}
    |\psi\rangle_4 = [\text{padding,} n_r,\text{padding,} \text{padding,}  K_T,\text{padding}]
\end{equation}

where each padding segment is initialised with zero amplitude and is of length $m/2$ where $m$ is the number of gridpoints of $n_r$ and $K_T$.

Finally, notice that for the earlier runs, we applied gaussian smoothing to the results obtained after each timestep before reinitialising.

\subsubsection{Noise and signal extraction}

We first tried to extract the QNMs frequencies by fitting the signals obtained from the unfiltered noisy simulations, and then  by applying a moving average smoothing method to i) only the space domain, ii) only the time domain, iii) both space and time domains (cf. fig.\ref{extract signal}, where to obtain a better signal to noise ratio we run each step twice setting the number of shots to the maximum available (8000)). From the earlier runs (cf. \ref{frequencies extracted plot} and table \ref{results table}), we observed that wave-like propagation in the signal, but notice that noise is also expected to propagate, making it difficult to fit the signal directly as we did in the noiseless qiskit classical simulation (cf. fig. \ref{QNMs convergence test}). Since we expect to find in the signal both the genuine QNMs frequencies and spurious frequencies due to noise, as an alternative to directly fitting the signal at a given radius, we obtain a global estimate and associated uncertainty using a non-parametric bootstrap method where we resample the fit obtained from each individual radial bin with replacement and and compute the mean frequency for each realization, first trimming each signal as to only include the segment where we expect the QNMs frequencies to be present, and smoothing it though a Savitzy-Golay filter.

Since performing the time derivative on the noisy signal would propagate the noise too, we use a nonlinear least-squares fit of the complex signal using the following analytically integrated damped sinusoid model
\begin{equation}
f(t)=\frac{A e^{-\omega_I t}}{\omega_R^2+\omega_I^2}\left[-\omega_I \cos \left(\omega_R t+\phi\right)+\omega_R \sin \left(\omega_R t+\phi\right)\right] 
\end{equation}
where $A$ is the amplitude, $\omega_R$ the oscillation frequency, $\omega_I$ the damping rate and $\phi$ the phase. The fit is carried out simultaneously on the real and imaginary parts of the signal with a small amplitude regularization term included to improve numerical stability. The initial values for the nonlinear least squares method are chosen using the following multi-mode heuristic ansatz: $A_0=\max |s(t)| / N_{\text {modes }}$, where $s(t)$ is the signal amplitude, $\omega_{R, n}^{(0)}=0.4 n$, $\omega_{I, n}^{(0)}=0.05 n$ and all the phases initialised as zero.
The final reported values correspond to the bootstrap mean, while the uncertainty was taken as the standard deviation of the bootstrap distribution. The values reported in fig.\ref{frequencies extracted plot} correspond to the bootstrap mean, while the uncertainty is the standard deviation of the bootstrap distribution. Frequencies are presented in the dimensionless form $M\omega$, where $M=1/2$ denotes the black-hole mass.

\section{Conclusion}
\label{conclusion}

With this project, we have developed a first proof-of-principle quantum algorithm for Numerical Relativity, using the Schwarzschild black hole quasinormal modes frequencies as a framework for validating the algorithm. The noiseless runs of the quantum circuit through Qiskit appear to be consistent with the expected behaviour and their agreement with results obtained from the classical finite difference simulation can be considered sufficient, for this proof-of-concept study, to support the validity and correct functioning of the quantum algorithm. We have also run the circuit components on physical IBM quantum backends through the UKRI National Quantum Computing Centre (NQCC) Quantum Access program  to determine how transpiling the circuit for physical backends would influence the runtime (fig. \ref{advection runtime_1}, \ref{collision runtime_2}, \ref{combinations runtime}). Finally, we have compared the quasinormal modes frequencies obtained from all runs (fig. \ref{frequencies extracted plot}). \\

Future improvements of the present framework may include the application of more advanced error mitigation techniques (e.g. zero noise extrapolation\cite{Giurgica_Tiron_2020}, probabilistic error correction \cite{gupta2023probabilisticerrorcancellationdynamic}, etc.), improved circuit optimization (eg. simplifying custom gates through their tensor networks representation \cite{Seitz_2023}) and implementing the alternative operators in the Appendix sections \ref{appendix implementing}, \ref{appendix collision}, \ref{appendix dynamical lapse}, \ref{appendix boundary}. 

From the runtime obtained from testing various parts of the circuit, we observe that the lapse subroutine is the most expensive one. We propose in the future to substitute it with the block encoding (\cite{Gily_n_2019, Berry_2014, Berry_2014_1, https://doi.org/10.4230/lipics.icalp.2019.33}) of the lapse function, that would be expected to scale polylogarithmically with the input size.\\

We considered particularly interesting the study of the Jacobian method to build the collision operator as an alternative to the exact reformulation of the nonlinear system through Carleman linearisation, since the related Hamiltonian can be easily modified by including variational parameters (e.g. rotation gates with tunable angles) that could be classically optimised (cf. VQE techniques) to test the phenomenology of beyond standard Relativity theories or to fit specific physics problems (e.g. strong gravity regimes). The spectrum of such a collision operator may be used to study the local relaxation rates of the system modes, invariants of the underlying dynamics and, if interpreted within the kinetic theory and lattice Boltzmann, entropy measures. In fact, it would be interesting to exploit the mathematical connections of the formalism we have employed with lattice-Boltzmann method (cf. section \ref{solving}) to translate definitions of equilibrium and entropy from the former to Relativity examples. In particular, it would be interesting to study the effect of the collision operator on the fields close to the black hole horizon. We conjecture that the collision operator decoheres the system around the horizon in such a way that coherences between field values across the horizon do not allow spurious of beyond classical Relativity effects such as echoes in quasinormal modes \cite{Hui_2019} or information transfer across the horizon.\\

This also suggests that interesting future applications may include both more complex classical physical systems ($1+1d \rightarrow 3+1d$) and problems at the intersection of gravitation and quantum physics . where this framework might provide an interesting laboratory for the study, as an example, of the information paradox\cite{raju2021lessonsinformationparadox} and of a quantum signatures in spacetimes (cf. e.g. \cite{Foo_2020, Foo_2022, Chakraborty_2025, coppo2026quantummodelblackholes, Giacomini_2021}). Furthermore, it would be interesting to study how a fuzzy horizon \cite{Dolan_2005} would change the frequencies of the extracted quasinormal modes.

However, it is important to note that before we can expect this framework to bear meaningful results in such exploratory areas of Physics, it will be necessary to improve the accuracy of the algorithm, as can be seen in our convergence tests todate. Fig. \ref{QNMs convergence test} shows the results obtained from Qiskit noiseless simulations run with 7 and 8 grid qubits compared with the classical, finite difference method simulation. While the limited convergence testing we have performed shows how the signal peaks move towards the classical solution as we increase the number of qubits (and hence the resolution, as we work with a discrete quantum walk with $\Delta t = \Delta r$), it is not yet possible to converge to the point where the framework can be applied to the more advanced examples mentioned above. At the time of writing, classically simulating or running on quantum backends circuits with a larger number of qubits would require computational costs beyond the intended scope of this study.\\

To conclude, even with the above mentioned limitations, we believe the presented framework may be of interest to both the Numerical Relativity and quantum computing research communities and might provide an interesting novel laboratory for the exploration of Physics at the intersection of those fields.

\section*{Acknowledgements}
The authors would like to thank Luisa T. Buchman for her continued support throughout this project, and Simon Williams, Eloy De Jong, Eugene Lim, Lionel London, Chiara Coviello, Yale Fan,  Matthew Duez and Marica Minucci for helpful discussions on various aspects of this work. We would also like to thank the organisers and participants to the University of Helsinki Cosmology Seminar (Helsinki, 25 Match 2026) and the workshop ``Concepts of Quantum and Spacetime" (KEK, Tsukuba, 9-12 March 2026), where part of this work was presented.

We acknowledge the use of the Qiskit software. This project was funded and supported by the UK National Quantum Computer Centre [NQCC200921], which is a UKRI Centre and part of the UK National Quantum Technologies Programme (NQTP).

The work of C.~Altomonte was supported by an NMES faculty grant. The work of M.~Fairbairn is supported by the U.K.\ Science and Technology Facilities Council (STFC) under Grant ST/X000753/1. 

\bibliographystyle{apsrev4-1}
\bibliography{biblio}

\appendix

\clearpage


\begin{center}
\textbf{\Large Appendix
}
\end{center}
\noindent

\section{Implementing $A_r \not= 0$}
\numberwithin{equation}{section}
\label{appendix implementing}

Consider the matrix
\begin{equation}
\boldsymbol{C}^{\hat{r}}=-\frac{1}{1-A_{\hat{r}}^2}\left(\begin{array}{cccc}
A_{\hat{r}} & 1 & 0 & 0 \\
1 & A_{\hat{r}} & 0 & 0 \\
0 & 0 & A_{\hat{r}} & 1 \\
0 & 0 & 1 & A_{\hat{r}}
\end{array}\right)
\end{equation}
For $A_r \not= 0$, we can rewrite each block part as 
\begin{equation}
M(A)=\left(\begin{array}{cc}
A & 1 \\
1 & A
\end{array}\right)=A I+X 
\end{equation}
then 
\begin{equation}
\boldsymbol{C}^{\hat{r}}=-\frac{1}{1-A_{\hat{r}}^2}\left[\left(A_{\hat{r}} I+X\right) \otimes I\right] 
\end{equation}
and hence
\begin{equation}
\boldsymbol{C}^{\hat{r}}=-\frac{A_{\hat{r}}}{1-A_{\hat{r}}^2} I \otimes I-\frac{1}{1-A_{\hat{r}}^2} X \otimes I 
\end{equation}
with eigenvalues 
\begin{equation}
\lambda_{ \pm}=-\frac{A_{\hat{r}} \pm 1}{1-A_{\hat{r}}^2}=-\frac{1}{1 \mp A_{\hat{r}}}
\end{equation}
Notice that $C^{\hat{r}}$ is not unitary and cannot be implemented directly as a unitary operator, but if $A_r$ varies with $r$, it is hermitian provided $A_{\hat{r}} \in \mathbb{R}$ and $A_{\hat{r}}^2 \neq 1$ everywhere and we can implement 
\begin{equation}
C^{\hat{r}}(r)=\alpha(r) I+\beta(r) X_1
\end{equation}
where
\begin{equation}
\alpha(r)=-\frac{A_{\hat{r}}(r)}{1-A_{\hat{r}}(r)^2}, \quad \beta(r)=-\frac{1}{1-A_{\hat{r}}(r)^2}
\end{equation}
and the related time evolution is 
\begin{equation}
U(r, t)=e^{-i t C^{\hat{r}}(r)}=e^{-i t \alpha(r)} e^{-i t \beta(r) X_1} 
\end{equation}
so we see that we can implement $e^{-i t \alpha(r)}$ in the same way as the static lapse encoding, i.e. as a set of phase gates controlled on the grid.

\section{Collision operator}
\label{appendix collision}

\subsection{Carleman linearisation}

When an exact algebraic representation is necessary, it is possible to use Carleman linearisation (or embedding)\cite{amini2022carlemanlinearizationnonlinearsystems, Akiba_Morii_Maruta_2023,Itani_Succi_2022}, which is a method for turning a nonlinear dynamical system into a larger linear system by introducing higher-order variables. Starting from 
\begin{equation}
\partial_t q+M \partial_r q=S(q), \quad q=\left(K_R, a_r, K_T, n_r\right)^{\top} 
\label{PDE and q}
\end{equation}
by promoting the nonlinear monomials to independent variables, and thus enlarging the state space 
\begin{equation}
q \mapsto Q=\left(q, q^{[2]}, q^{[3]}, \ldots\right)
\end{equation}
where $q^{[n]}=q^{\otimes n}$ is the degree- $n$ tensor monomials (Kronecker powers) of the fields, then the nonlinear PDE becomes a larger (formally infinite, it can then be truncated at a given maximum degree of the tensor monomials) linearised system
\begin{equation}
    \partial_t \mathcal{Q}+\mathcal{M} \partial_r \mathcal{Q}=\mathcal{A} \mathcal{Q}
\end{equation}
where $\mathcal{M}$ and $\mathcal{A}$ are matrices acting on the expanded Hilbert space.

\subsection{Local linearisation through Jacobian matrix} 

\begin{figure*}
    \centering
    \includegraphics[width=0.8\linewidth]{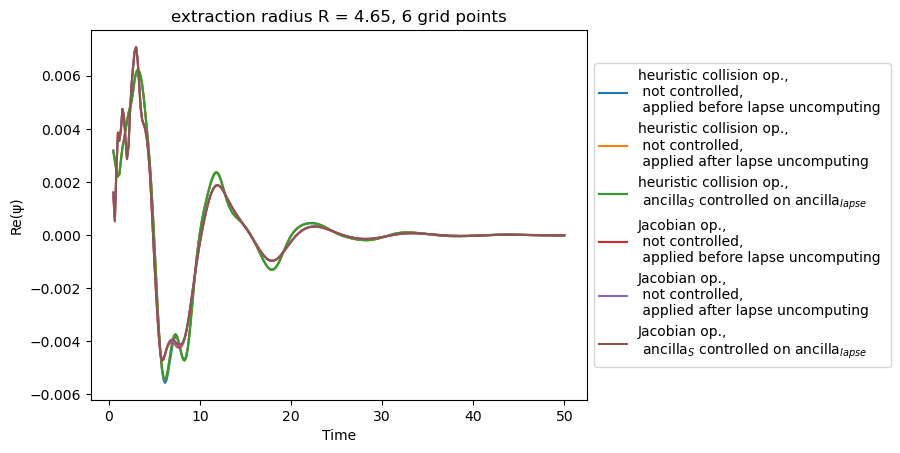}
    \caption{\it Comparison between Jacobian-derived and heuristic collision operators with different methods to control the $S$ operator on ancilla$_{lapse}$, from noiseless Qiskit simulations with 6 grid qubits.}
    \label{jacobian vs heuristic}
\end{figure*}

Taking inspiration from linearised lattice-Boltzmann methods, where a nonlinear collision term is approximated as a local linear operator through its Jacobian \cite{Junk_Yong_2009,Yang_Zhao_Lin_2024}, the source term in eq.\ref{PDE and q} can be approximated locally as 
\begin{equation}
    S(q) \approx S\left(q_0\right)+J\left(q_0\right)\left(q-q_0\right)
    \label{jacobian approx}
\end{equation}
where $q \equiv q\left(x, t\right)$, $q_0 \equiv q\left(x_0, t\right)$, and $J_{i j}=\frac{\partial S_i}{\partial q_j}$ is the Jacobian matrix. By defining $\delta q=q-q_0$ and $\delta S=S(q)-S\left(q_0\right)$, the linearised local dynamics is 
\begin{equation}
\delta S \approx J(q_0)  \delta q
\end{equation}
Notice that eq.\ref{jacobian approx} can be connected to the collision (or relaxation) step of the Lattice-Boltzmann method \cite{Bellotti_Helluy_Navoret_2025, Fei_2017})
\begin{equation}
    f^*=f+\Delta t J(q) f
\end{equation}
where $f \quad \leftrightarrow \quad \delta q=q-q_0$, so that the linearised dynamics become $\Delta f \sim J f$.\\

\subsubsection{Schwarzschild example}

Hence, our aim is to implement the local operator $J$ for our specific example by means of a quantum circuit component. We suggest it can be done in the following way. First write the Jacobian matrix
\begin{equation}
\begin{aligned}
\mathbf{J}_{\mathbf{f}}(K_R,a_r,K_T,n_r)
&=
\left[
\begin{array}{llll}
\frac{\partial f_1}{\partial K_R} &
\frac{\partial f_1}{\partial a_r} &
\frac{\partial f_1}{\partial K_T} &
\frac{\partial f_1}{\partial n_r} \\
\frac{\partial f_2}{\partial K_R} &
\frac{\partial f_2}{\partial a_r} &
\frac{\partial f_2}{\partial K_T} &
\frac{\partial f_2}{\partial n_r} \\
\frac{\partial f_3}{\partial K_R} &
\frac{\partial f_3}{\partial a_r} &
\frac{\partial f_3}{\partial K_T} &
\frac{\partial f_3}{\partial n_r} \\
\frac{\partial f_4}{\partial K_R} &
\frac{\partial f_4}{\partial a_r} &
\frac{\partial f_4}{\partial K_T} &
\frac{\partial f_4}{\partial n_r}
\end{array}
\right]
\\[1ex]
&\hspace{-1cm}=
\left[
\begin{array}{cccc}
-2 K_R & 2 a_r & 2 K_T & -2 n_r \\
0 & 0 & 0 & 0\\
K_T & -n_r & K_R-2K_T & -2 n_r \\
n_r & -K_T & -a_r-2 n_r & K_R-2 K_T 
\end{array}
\right]
\\[1ex]
&\hspace{-1cm}+
\left[
\begin{array}{cccc}
0 & 0 & 0 & 2/R \\
2/R & 0 & -2/R & 0\\
0 & 1/R & 0 & 1/R \\
-1/R & 0 & 1/R & 0 
\end{array}
\right]
\end{aligned}
\end{equation}
where we can separate the two contributions 
\begin{equation}
J=J_{\text {nonlinear }}+\frac{1}{R} J_{\text {geom }}
\end{equation}
from which we want to build an effective collision operator of the form 
\begin{equation}
J_{op}=\sum_{i, j} a_{i j}|i\rangle\langle j|+\frac{1}{R} \sum_{i, j} b_{i j}|i\rangle\langle j|
\end{equation}
that we will later unitary-embedded in an operator 
\begin{equation}
H_{J_{op}}=\left(\begin{array}{cc}
0 & J_{op} \\
J_{op}^{\dagger} & 0
\end{array}\right)
\end{equation}
and implement in the quantum circuit as a Hamiltonian gate $U_{J_{op}}(\Delta  t)=e^{-i \Delta t H_{J_{op}}}$. \\

Notice that $J_{\text {nonlinear }}$ encodes the nonlinear part of $S$, while $J_{\text {geom }}$ encodes the part of $S$ that is linear in the fields, but contains the dependence on the geometry. Because of this, it is preferable to obtain the generators $H_{\text {nonlinear }}$ and $H_{\text {geom }}$ separetely, and apply the operator splitting (analogous to Lie-Trotter step)
\begin{equation}
U(\Delta t) \approx e^{-i \Delta t H_{\text {nonlinear }}} e^{-i \Delta t \frac{1}{R} H_{\text {geom }}}
\end{equation}
(or higher order) which is a good approximation in the limit where $\Delta t\left\|\left[H_{\text {nonlinear }}, H_{\text {geom }}\right]\right\| \ll 1$. Below we comment on how to obtain the generators.\\

Consider first $J_{\text {nonlinear }}$. We model it as an effective collision operator built from $J_{\text {nonlinear }}$ as follows: we interpret $|i\rangle$ as indicating the field in a given $c_{ij}$ of $J_{\text {nonlinear }}$, $a_{i j}$ as the coefficient multiplying it, and $|j\rangle$ as the row of $J_{\text {nonlinear }}$, i.e.
\begin{equation}
\sum_{i, j} a_{i j}|i\rangle\langle j| = \sum_{ij} \delta_{i, j} \times \text{field coeff}(c_{ij})|i\rangle\langle j|
\end{equation}
which means that amplitude of field $i$ is locally transferred to field $j$ according to intensity proportional to value of matrix entry. Similarly we can populate all of $\sum_{i, j} a_{i j}|i\rangle\langle j|$. As an example, from the first row of $J_{\text {nonlinear }}$ we get the following matrix elements 
\begin{equation}
    \sum_i\left(-2 \delta_{i, K_R}+2 \delta_{i, a_r}+2 \delta_{i, K_T}-2 \delta_{i, n_r}\right)\left|i\right\rangle\langle K_R|
\end{equation}
Regarding $J_{\text {geom }}$,  as mentioned it encodes the part of $S$ that is linear in the fields, but contains the dependence on the geometry. Its effect can be interpreted as multiplying the fields by $1/R$ and then redistributing such amplitudes. Hence, we can readily read off the components from $J_{\text {geom }}$ as follows: 
\begin{equation}
\frac{1}{R} \sum_{i, j} b_{i j}|i\rangle\langle j| = \frac{1}{R}\sum_{ij} c_{ij} |i\rangle\langle j|
\end{equation}

\subsubsection{Simplification for the Schwarzschild example}

Finally, notice that, for the example at hand, 
\begin{equation}
n_{\hat{r}}=\frac{1-D_{\hat{r}} R}{R} 
\end{equation}
which rearranged gives 
\begin{equation}
\frac{1}{R}=n_r+D_{\hat{r}}(\ln R)
\end{equation}
where $D_{\hat{r}}(\ln R)$ can be interpreted as a coefficient dependent on the background geometry. Hence, in the limit where such a background coefficient varies slowly over the interaction scale of the local effective collision operator $J_{o p}$, we can set $\frac{1}{R}\approx n_r$ and absorb $\frac{1}{R} J_{\text {geom }}$ into $J_{\text {nonlinear }}$.

\subsection{Heuristic collision operator}

Taking inspiration from heuristic kinetic/multi-relaxation time lattice-boltzmann methods\cite{osti_447041, Chalabi_Qiu_2000, Pareschi_2001, krishnamurthy2013variablerelaxedschemesmultidimensional, Crin-Barat_Shou_2023, dHumieres_2002, Zhang_Zhao_Lin_2019}, we can interpret the source $S$ as acting as an heuristic relaxation operator, or a mode-mixing mechanism. This is close to a graph/ network interpretation of the source 
\begin{equation}
\dot{q}_i=\sum_j \Omega_{i j}(q, R) q_j 
\end{equation}
so that the collision/source step is
\begin{equation}
f_i^{a, *}=f_i^a+\Delta t \sum_b \Omega_{a b}(q, R)\left(f_i^b-f_i^a\right)
\end{equation}
Hence, we aim at finding a generalised collision matrix, $\Omega_{a b}$ that encodes the \textit{interaction topology} of the fields, e.g. 

$$
K_R \leftrightarrow n_{\hat{r}}, \quad K_T \leftrightarrow a_{\hat{r}}, \quad \text { etc. }
$$

which works heuristically and check whether it is a physically faithful representation of the mixing of field populations across channels, dissipation mechanisms, qualitative nonlinear feedback, and if it preserves the stability of the scheme.\\

To build it, we identify dominant couplings and build
approximate relaxation operators as follows.

From 
\begin{equation}
\begin{aligned}
& S_{-} K_R=a_{\hat{r}}^2-n_{\hat{r}}^2-K_R^2+K_T^2+\frac{2 n_{\hat{r}}}{R} \\
& S_{-} a_{\hat{r}}=\frac{2\left(K_R-K_T\right)}{R}\\
& S_{-} K_T=\frac{a_{\hat{r}}+n_{\hat{r}}}{R}+K_R K_T-K_T^2-n_{\hat{r}}^2-a_{\hat{r}} n_{\hat{r}}, \\
& S_{-} n_{\hat{r}}=\frac{K_T-K_R}{R}-a_{\hat{r}} K_T+n_{\hat{r}}\left(K_R-2 K_T\right) 
\end{aligned}
\end{equation}
build ansatz coupling matrix
\begin{equation}
\Omega_1 =\left[\begin{array}{cccc}
0 & 0 & -1 & 0 \\
0 & 0 & 0 & -1 \\
-1 & 0 & 0 & 0 \\
0 & -1 & -2 & 0
\end{array}\right] \quad \begin{aligned}
& \rightarrow \propto K_T K_R \text{coupling} \\
& \rightarrow \propto n_r a_r \text{coupling} \\
& \rightarrow\propto K_R K_T \text{coupling} \\
& \rightarrow\propto a_r n_r,  K_T n_r \text{coupling} 
\end{aligned}
\end{equation}
and $R$-dependent matrix
\begin{equation}
\Omega_2 =\left[\begin{array}{cccc}
0 & 0 & 0 & 2 / R \\
2 / R & 0 & 1 / R & 0 \\
1 / R & -1 / R & 0 & 0 \\
-1 / R & 1 / R & 0 & 0
\end{array}\right]
\begin{aligned}
& \rightarrow \propto n_r K_R \text{ coupling} \\
& \rightarrow \propto K_R a_r, K_T a_r \text{ coupling} \\
& \rightarrow\propto K_R K_T, a_rK_T \text{ coupling} \\
& \rightarrow\propto K_R n_r,  a_r n_r \text{ coupling} 
\end{aligned}
\end{equation}
and then 
\begin{equation}
    \Omega_{ij} = \Omega_{1,ij} + \Omega_{1,ij}
\end{equation}
that we will later unitary-embedded in an operator

\begin{equation}
H_{\Omega}=\left(\begin{array}{cc}
0 & \Omega \\
\Omega^{\dagger} & 0
\end{array}\right)
\end{equation}
and implement in the quantum circuit as a Hamiltonian gate $U_{\Omega}(\Delta t)=e^{-i \Delta t H_{\Omega}}$.

\section{Dynamical lapse and shift subroutines}
\label{appendix dynamical lapse}

\subsection{Dynamical lapse}

\subsubsection{encoding}
 
To obtain a dynamical lapse function, we want to encode in an ancilla the solution to eq.(\ref{lapse function}),
\begin{equation}
\alpha(r)=\alpha_0 \exp \left(\int_{r_0}^{r_{m-1}} a_{\hat{r}}(r') d r'\right) 
\end{equation}
which can be approximated as a  cumulative sum $\alpha_m=\alpha_0 e^{S_m}$ where $S_m=\sum_{i=0}^{m-1} a_i \Delta r_i$.

Our strategy is to encode a function of $a_{\hat{r}}$ in the amplitude of an ancilla qubit, then perform the cumulative sum, (optionally exponentiate the amplitude), and control the required operations on the ancilla qubit, then disentangle the ancilla qubit from the coin and grid qubits. \\

For every $r$, apply $R_y\left(\theta\right)$ to the ancilla qubit $q_a$, conditioned on the coin qubits selecting the binary combination corresponding to $a_{\hat{r}}$ (for our encoding, $|00\rangle_c$), and the grid qubits selecting the binary combination corresponding to $r$, to obtain
\begin{equation}
\begin{aligned}
\sum_r \psi_{00, r} |00\rangle_c | r\rangle \left(\cos \frac{\theta}{2} |0\rangle_a+\sin \frac{\theta}{2} |1\rangle_a\right)\\
+\sum_{c \neq 00, r} \psi_{c, r} |c\rangle |r\rangle |0\rangle_a 
\end{aligned}
\end{equation}
Then, under the small angle approximation 
\begin{equation}
\Psi \approx \frac{\theta}{2} \sum_r \psi_{00, r} |00\rangle_c |r\rangle |1\rangle_a + \sum_{c, r} \psi_{c, r} |c\rangle |r\rangle |0\rangle_a
\end{equation}
So, the amplitudes of the $|1\rangle_a$ branch of the ancilla now contains the \textit{linear amplitude encoding} of $a_{\hat{r}}$.

\subsubsection{integration}

A coherent adder operator maps amplitudes $f_j$ to cumulative sums 
\begin{equation}
g_j=\sum_{k<j} f_k 
\end{equation}
can be implemented as a linear operator $A$ acting on a $2^m$ dimensional grid register, where
\begin{equation}
A_{j k}= \begin{cases}1, & k<j \\ 0, & k \geq j\end{cases}
\end{equation}
e.g. for the $2^2$ grid example in fig.\ref{alg:dynamic lapse}, 
\begin{equation}
A=\left(\begin{array}{llll}
0 & 0 & 0 & 0 \\
1 & 0 & 0 & 0 \\
1 & 1 & 0 & 0 \\
1 & 1 & 1 & 0
\end{array}\right) 
\end{equation}
which, acting on a vector $\psi_r=\left(\psi_{r_0}, \psi_{r_1}, \psi_{r_2}, \psi_{r_3}\right)^T$, gives 
\begin{equation}
A \psi_r=\left(\begin{array}{c}
0 \\
\psi_{r_0} \\
\psi_{r_0}+\psi_{r_1} \\
\psi_{r_0}+\psi_{r_1}+\psi_{r_2}
\end{array}\right) 
\end{equation}
To apply it to $\sum_r \psi_{00, r}|00\rangle_c|r\rangle|1\rangle_a$, we embed $A$ in the block operator 
\begin{equation}
H=\left(\begin{array}{cc}
0 & A \\
A^{\dagger} & 0
\end{array}\right)
\end{equation}
which acts as
\begin{equation}
H|\Psi\rangle=\frac{\theta}{2}\sum_r(A \psi_{00,r})|00\rangle_c|r\rangle|0\rangle_a + \sum_{c,r}(A^{\dagger} \phi_c)|c\rangle|r\rangle|1\rangle_a
\end{equation}
but to implement it in a quantum circuit, we need to apply $e^{-i H t}$ to the ancilla qubit and grid register, which in the small time limit is $e^{-i t H} \approx I-i t H$.\\

Notice at this point the $|0\rangle_a$ branch carries a discretised version of the integral, since 
\begin{equation}
(A \psi)_m \sim \sum_{k<m} \psi\left(r_k\right) \approx \int_{r_0}^{r_m} \psi\left(r^{\prime}\right) d r^{\prime}
\end{equation}
To obtain the exponential, we can use phase-kickback on an additional ancilla controlled on $|0\rangle_a$

\subsubsection{Conditioning operations on ancilla qubit and uncomputing amplitude encoding}

As in section \ref{fixed lapse section}, we control the required operations on the ancilla qubit and then disentangle the ancilla qubit by applying $U^{\dagger}$, where $U$ corresponds to the dynamical lapse encoding.

\begin{algorithm}[H]
\caption{Dynamical lapse}
\label{alg:dynamic shift}
\begin{algorithmic}[1]
\Require 1 ancilla qubit for encoding
\For {$i = 0$ to $m$}
    \Statex rotate ancilla using $RY(\theta)$ (small $\theta$) controlled on coin qubits boolean value $(00) = a_r$ and grid qubits corresponding to grid position $x_i$;
\EndFor
\State Apply integrator to ancilla and grid through $e^{-iHt}$;
\State Control required operations on ancilla value 1;
\State Undo integration step through applying $e^{+iHt}$ to ancilla and grid;
\For {$i = 0$ to $m$}
    \Statex rotate ancilla using $RY(\theta)$ (small $\theta$) controlled on coin qubits boolean value $(00) = a_r$ and grid qubits corresponding to grid position $x_i$.
\EndFor
\end{algorithmic}
\end{algorithm}

\subsection{Dynamical shift}

We can obtain a dynamical shift through the same mechanism as in the previous secction, with $R_y(\theta)$ now applied to the ancilla qubit $q_b$, and the controls on the coin register now selecting $K_R$
\begin{equation}
\sum_r \psi_{01, r}|01\rangle_c|r\rangle\left(\cos \frac{\theta}{2}|0\rangle_b+\sin \frac{\theta}{2}|1\rangle_b\right)+\sum_{c \neq 0, r} \psi_{c, r}|c\rangle|r\rangle|0\rangle_b
\end{equation}
Notice that we also need to condition on $q_a = |1\rangle_a$ to  obtain an amplitude encoding on $q_b = |1\rangle_b$ $\propto f(\alpha)K_R$.\\

\begin{algorithm}[H]
\caption{Dynamical shift}
\label{alg:dynamic_lapse}

\begin{algorithmic}[1]

\Require 1 ancilla qubit for encoding

\For{$i = 0$ to $m$}
    \Statex rotate ancilla using $RY(\theta)$ (small $\theta$) controlled on coin qubits boolean value $(01)$ corresponding to $K_T$ and grid qubits corresponding to grid position $x_i$;
\EndFor

\State Apply integrator to ancilla and grid through $e^{-iHt}$;

\State Control required operations on ancilla value 1;

\State Undo integration step through applying $e^{+iHt}$ to ancilla and grid;

\For{$i = 0$ to $m$}
    \Statex rotate ancilla using $RY(\theta)$ (small $\theta$) controlled on coin qubits boolean value $(01)$ corresponding to $K_T$ and grid qubits corresponding to grid position $x_i$.
\EndFor
\end{algorithmic}
\end{algorithm}

\begin{figure}
    \centering
    \includegraphics[width=1\linewidth]{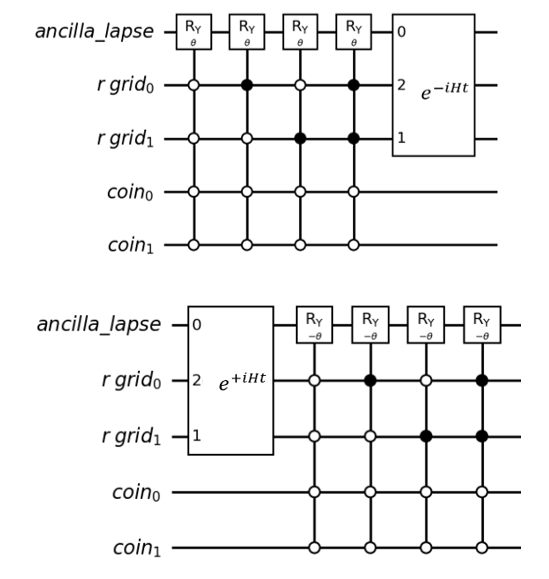}
    \caption{\it Dynamical lapse encoding and disentangling}
    \label{fig:placeholder}
\end{figure}

\section{Boundary conditions: amplitude leakage operator}
\label{appendix boundary}

First, we double the grid domain with an ancilla represeting an auxiliary enviroment.  
\begin{equation}
|\psi\rangle_S \otimes|0\rangle_E
\end{equation}
and define a \textit{physical sector} $|i\rangle_S|0\rangle_E \quad$ and an \textit{absorbed sector} $|i\rangle_S|1\rangle_E \quad$. Note in the circuit we add the ancilla on top of the grid register, which, following the little-endian convention, corresponds to the convention where the final boolean value corresponds to the ancilla, i.e. $|i, b\rangle \leftrightarrow 2 i+b$ (also known as \textit{interleaved tensor product encoding}, or \textit{binary ancilla embedding}). The structure of the operator in eq. (\ref{boundary operator}) reflects this convention.\\

Then, we define a system-environment unitary acting locally at selected boundary indices in such a way that the amplitude at the selected system grid locations is leaked to the environment. As an example, we can implement, At each field's boundary site $i \in \mathcal{B}$, 
\begin{equation}
\begin{aligned}
|i\rangle_S|0\rangle_E & \mapsto \sqrt{1-p}|i\rangle_S|0\rangle_E+\sqrt{p}|i\rangle_S|1\rangle_E \\
|i\rangle_S|1\rangle_E & \mapsto \text { (unitary completion) }
\end{aligned}
\end{equation}
and finally trace out environment. This effectively takes amplitude in system mode $|i\rangle_S$, splits it into two branches, one which stays in the system with weight $1-p$ and one that goes to the ancilla with weight $p$. Finally, by tracing out the ancilla
\begin{equation}
\rho_S^{\prime}=\operatorname{Tr}_E\left(\rho_{S E}^{\prime}\right)
\end{equation}
the system loses probability mass at those sites. For example, we can build the following unitary operator 
\begin{equation}
U=\bigoplus_{i \notin \mathcal{B}} I_2 \oplus \bigoplus_{i \in \mathcal{B}}\left(\begin{array}{cc}
\sqrt{p} & \sqrt{1-p} \\
\sqrt{1-p} & -\sqrt{p}
\end{array}\right) 
\label{boundary operator}
\end{equation}
acting on $\{|i, 0\rangle,|i, 1\rangle\}$ with ancilla initially in $|0\rangle$.

\begin{figure}
    \centering
    \includegraphics[width=0.6\linewidth]{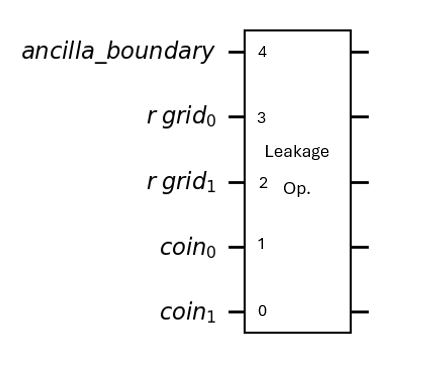}
    \caption{\it Leakage operator. Notice the ordering of qubits to which it is applied to respect the tensor structure in the described construction.}
    \label{fig:placeholder}
\end{figure}

\section{QNMs}
\label{appendix qnms}

\subsection{Perturbation of a spherically symmetric spacetime: theory summary}

The WEBB Schwarzschild metric (eq.\ref{WEBB metric}) is written in such a way that it can readily be related to the metric for a spherically symmetric background used in \cite{Sarbach_2001} (eq. 1 therein) and \cite{Buchman_2007} (eq. 5 therein) for the study of perturbations of such a spacetime. Hence, in this section we first summarise the relevant equations from \cite{Sarbach_2001, Buchman_2007}, and show how we apply them to the problem at hand.\\

\subsubsection{Summary of relevant quantities}

It is useful to first write the metric in the following $2+2$ form $\tilde{M} \times S^2$ 
\begin{equation}
ds^2=\tilde{g}_{a b} d x^a d x^b+r^2 \hat{g}_{A B} d x^A d x^B
\label{WEBB metric_1}
\end{equation}
where lower-case letters $a,b=\{t,r\}$ can take the value of time and radial quantities, while capital letters $A,B=\{\theta,\phi\}$  correspond to angular quantities.
Hence, we see that $\hat{g}_{A B}$ is the standard metric on $S^2$ (with $\hat{g}_{\phi\phi}=\sin^2\theta$), and $\tilde{g}$ is the metric on the two-dimensional pseudo-Riemannian orbit space $\tilde{M}$.\\

Since the framework of numerical relativity is used in both \cite{Buchman_2005} and \cite{Sarbach_2001}, it is useful to also recall how eq. (\ref{WEBB metric_1}) above can be written in ADM form
\begin{equation}
ds^2 =-\alpha(\mu)^2 d t^2+\bar{g}_{i j}(\mu)\left(d x^i+\beta^i(\mu) d t\right)\left(d x^j+\beta^j(\mu) d t\right)
\end{equation}
where $\mu$ is a variational parameter, such that for $\mu=0$, the metric is spherically symmetric, $\bar{g}_{i j}$ is the the spatial metric on the 3-dimensional hypersurfaces. This ADM metric can also be written in the $2 + 2$ form $\tilde{M} \times S^2$ described above (\ref{WEBB metric_1}), with the orbit metric (the part only containing $t,r$) is given by 
\begin{equation}
\tilde{g_{ab}}dx^adx^b=-\alpha^2 d t^2+\gamma^2(d x+\beta d t)^2
\end{equation}
where $x$ is any radial coordinate, $\alpha$ and $\beta \equiv \beta^x$ are the background lapse and shift, respectively, and $\gamma^2 \equiv \bar{g}_{x x}$.\\

To study Schwarzschild black hole perturbations using the framework in \cite{Sarbach_2001}, the components of the linearized metric and extrinsic curvature are needed.\\

The extrinsic curvature components are given by 
\begin{equation}
\begin{aligned}
& 2 \alpha K_{x x}=2 \gamma\left(\partial_0 \gamma-\gamma \beta^{\prime}\right) \\
& 2 \alpha K_{x A}=0 \\
& 2 \alpha K_{A B}=2 r \partial_0 r \hat{g}_{A B}
\end{aligned}
\end{equation}
where $x$ stands for any radial coordinate, and $A,B \in \{\theta,\phi\}$. The linearised metric components are given by 
\begin{equation}
\begin{aligned}
\delta g_{t t} & =-2 \alpha \delta \alpha-\beta^2 \delta \bar{g}_{x x}+2 \beta \delta \beta_x \\
\delta g_{t j} & =\delta \beta_j \\
\delta g_{i j} & =\delta \bar{g}_{i j}
\end{aligned}
\end{equation}

\subsubsection{QNMs}
The metric and extrinsic curvature perturbations supporting the QNMs of the odd-parity sector with $l \geqslant 2$ are 
\begin{equation}
\begin{aligned}
& \delta \beta_A=b S_A \\
& \delta \bar{g}_{x A}=h_1 S_A, \quad \delta \bar{g}_{A B}=2 h_2 \hat{\nabla}_{(A} S_{B)} \\
& \delta K_{x A}=\pi_1 S_A, \quad \delta K_{A B}=2 \pi_2 \hat{\nabla}_{(A} S_{B)} 
\end{aligned}
\end{equation}
Then the Regge-Wheeler function is given by 
\begin{equation}
\Phi=\frac{r}{\lambda \alpha \gamma}\left(2 \alpha \pi_1-2 \frac{\partial_0 r}{r} h_1\right)
\label{PHI}
\end{equation}
where 
\begin{equation}
\begin{aligned}
&2 \alpha \pi_1=\dot{h}_1-2 \beta \frac{r^{\prime}}{r} h_1-r^2\left(\frac{b}{r^2}\right)^{\prime}\\
&2 \alpha \pi_2=\partial_0 h_2+\beta h_1-b 
\label{pi1 pi2}
\end{aligned}
\end{equation}
where a dot denotes $\partial_t$, a prime denotes differentiation with respect to $x$, and $\partial_0 \equiv \partial_t-\beta \partial_x$. Eqs.(\ref{pi1 pi2}) can be rearranged into gauge invariant quantities
\begin{equation}
\begin{aligned}
& h_0^{(i n v)}=-2 \alpha \pi_2+2 \frac{\partial_0 r}{r} h_2 \\
& h_x^{(i n v)}=h_1-r^2\left(\frac{h_2}{r^2}\right)^{\prime}
\end{aligned}
\end{equation}
related to $\Phi$ as
\begin{equation}
\begin{aligned}
&h_0^{(i n v)}=-\frac{\alpha}{\gamma} \partial_x(r \Phi)\\
&h_x^{(i n v)}=-\frac{\gamma}{\alpha} \partial_0(r \Phi) 
\end{aligned}
\end{equation}

\subsection{Strategy}

Here we consider arguments as to why a suitable perturbation of the WEBB Schwarzschild metric in eq.(\ref{WEBB metric}) can be projected onto the odd parity sector of gravitational QNMs. Before giving the full details in the following sections, we provide here a summary of the arguments:

\begin{enumerate}
    \item Argument 1 summary: metric (eq.\ref{WEBB metric}) supports $\{\delta \bar{g}_{A B}, \delta K_{A B}\} \neq 0$, and hence $\{h_2,\pi_2\} \neq 0$. While $\{\delta \bar{g}_{x A}, \delta K_{x A}\} = 0$ and hence $\{h_1, \pi_1 \} = 0$ initially, eqs. (\ref{pi1 pi2}) relate $\{h_2, \pi_2 \}$ and $\{h_1, \pi_1 \}$. Hence, provided initial perturbations related to $\{h_2, \pi_2 \}$ are present, it should be possible to have $\{h_1, \pi_1 \} \neq 0$ and hence $\Phi \neq 0$.

    \item Argument 2 summary: we write the Newman-Penrose scalar $\Psi_4$ as a function of Ricci rotation coefficients, and show how a perturbation of the background metric though $\delta n_r$ and $\delta K_r$ is expected to modulate the QNMs frequencies.
\end{enumerate}

\subsection{Argument 1}

To relate this to our example, notice that $\hat{\nabla}_{(A} S_{B)}$ is a symmetric rank-2 tensor on the 2-sphere. Note that for both $A \neq B$ and $A=B$, this it is generally non-zero (in particular, for $A=B$, while tensor harmonic is tracefree, individual diagonal components are non-zero).\\

It follows that for $A=B=\theta, \phi, \delta \bar{g}_{A B}$ and $\delta K_{A B}$ are in general $\neq 0$. Looking at(\ref{WEBB metric}), we have 
\begin{equation}
  \delta \bar{g}_{\theta \theta} \propto \delta R \propto \delta n_r \propto h_2
\end{equation}
\begin{equation}
 \delta \bar{g}_{\phi \phi} \propto \sin \theta \delta R \propto \sin \theta \delta n_r \propto h_2
\end{equation}
and considering the Ricci rotation coefficients related to the esxtrinsic curvature present in our example, we have 
\begin{equation}
\delta K_{\theta \theta} = \delta K_{\hat{\theta} \hat{\theta}}=\delta K_{\hat{\phi} \hat{\phi}} = \delta K_T \propto \pi_2
\end{equation}
While perturbations 
\begin{equation}
\begin{aligned}
& \delta \bar{g}_{x A}=h_1 S_A \\
& \delta K_{x A}=\pi_1 S_A
\end{aligned}
\end{equation}
are originally absent from the metric, from eq. (\ref{pi1 pi2}), we see than if $h_2, \pi_2$ are present, they can induce $\neq h_1, \pi_1$, which enter eq.(\ref{PHI}) 

\subsection{Argument 2: Newman-Penrose formalism}

In the Newman-Penrose formalism, the Weyl scalar $\Psi_4$ is the component of the Weyl tensor projected onto the Newman-Penrose null tetrad basis $\left\{l^a, n^a, m^a, \bar{m}^a\right\}$
\begin{equation}
\Psi_4 \equiv-\delta C_{a b c d} n^a \bar{m}^b n^c \bar{m}^d
\label{np}
\end{equation}
that encodes the outgoing gravitational radiation. For waves propagating in the radial direction in the transverse-traceless gauge, it is equivalent to
\begin{equation}
\Psi_4=\frac{1}{2}\left(\ddot{h}_{\hat{\theta} \hat{\theta}}-\ddot{h}_{\hat{\phi} \hat{\phi}}\right)+i \ddot{h}_{\hat{\theta} \hat{\phi}}=-\ddot{h}_{+}+i \ddot{h}_{\times}
\end{equation}\\

For vacuum perturbations, $\delta C_{a b c d}=\delta R_{a b c d}$, so we can compute eq.\ref{np} in terms of Riemann tensor components, where 
\begin{equation}
\begin{split}
R_{\alpha \beta \gamma \delta} &=
D_\gamma \Gamma_{\alpha \beta \delta} - D_\delta \Gamma_{\alpha \beta \gamma} \\
&\quad + \Gamma_{\alpha \varepsilon \gamma} \Gamma_{\beta \delta}^\epsilon
- \Gamma_{\alpha \varepsilon \delta} \Gamma_{\beta \gamma}^{\varepsilon} \\
&\quad + \Gamma_{\alpha \beta \varepsilon}
\left(\Gamma_{\gamma \delta}^{\varepsilon} - \Gamma_{\delta \gamma}^{\varepsilon}\right)
\end{split}
\end{equation}
Starting from the WEBB orthonormal tetrad in the toy model $\left(e_{\hat{t}}, e_{\hat{r}}, e_{\hat{\theta}}, e_{\hat{\phi}}\right)$, define

\begin{equation}
\begin{aligned}
l^\mu & =\frac{1}{\sqrt{2}}\left(e_{\hat{t}}^\mu+e_{\hat{r}}^\mu\right) \\
n^\mu & =\frac{1}{\sqrt{2}}\left(e_{\hat{t}}^\mu-e_{\hat{r}}^\mu\right) \\
m^\mu & =\frac{1}{\sqrt{2}}\left(e_{\hat{\theta}}^\mu+i e_{\hat{\phi}}^\mu\right)
\end{aligned}
\end{equation}

Then, eq.(\ref{np}) can be expanded as
\begin{equation}
\begin{aligned}
\Psi_4 &=
-\frac12
\Big[
\delta R_{\hat t\hat\theta\hat t\hat\theta}
-\delta R_{\hat t\hat\phi\hat t\hat\phi}
-\delta R_{\hat r\hat\theta\hat r\hat\theta}
+\delta R_{\hat r\hat\phi\hat r\hat\phi}
\Big] \\
&\quad
+ i
\Big[
\delta R_{\hat t\hat\theta\hat t\hat\phi}
-\delta R_{\hat r\hat\theta\hat r\hat\phi}
\Big].
\end{aligned}
\label{psi_4 expanded}
\end{equation}
Let 
\begin{equation}
\Gamma_{\alpha \beta \gamma}=\bar{\Gamma}_{\alpha \beta \gamma}+\delta \Gamma_{\alpha \beta \gamma} 
\end{equation}
then 
\begin{equation}
R_{\alpha \beta \gamma \delta}=\bar{R}_{\alpha \beta \gamma \delta}+\delta R_{\alpha \beta \gamma \delta}+O\left(\delta \Gamma^2\right)
\end{equation}
with 
\begin{equation}
\begin{aligned}
\delta R_{\alpha \beta \gamma \delta}= & \bar{D}_\gamma \delta \Gamma_{\alpha \beta \delta}-\bar{D}_\delta \delta \Gamma_{\alpha \beta \gamma} \\
& +\bar{\Gamma}_{\alpha \epsilon \gamma} \delta \Gamma_{\beta \delta}^\epsilon+\delta \Gamma_{\alpha \epsilon \gamma} \bar{\Gamma}_{\beta \delta}^\epsilon \\
& -\bar{\Gamma}_{\alpha \epsilon \delta} \delta \Gamma_{\beta \gamma}^\epsilon-\delta \Gamma_{\alpha \epsilon \delta} \bar{\Gamma}_{\beta \gamma}^\epsilon 
\end{aligned}
\end{equation}
which is the linearised curvature around an arbitrary background tetrad. Then, the addends of the real part of eq.(\ref{psi_4 expanded}) are
\begin{equation}
\begin{gathered}
\delta R_{t \theta t \theta}=\bar{D}_t \delta \Gamma_{t \theta \theta}-\bar{D}_\theta \delta \Gamma_{t \theta t} \\
+\bar{\Gamma}_{t \epsilon t} \delta \Gamma_{\theta \theta}^\epsilon+\delta \Gamma_{t \epsilon t} \bar{\Gamma}_{\theta \theta}^\epsilon \\
-\bar{\Gamma}_{t \epsilon \theta} \delta \Gamma_{\theta t}^\epsilon-\delta \Gamma_{t \epsilon \theta} \bar{\Gamma}_{\theta t}^\epsilon 
\end{gathered}
\end{equation}
\begin{equation}
\begin{gathered}
\delta R_{t \phi t \phi}=\bar{D}_t \delta \Gamma_{t \phi \phi}-\bar{D}_\phi \delta \Gamma_{t \phi t} \\
+\bar{\Gamma}_{t \epsilon t} \delta \Gamma_{\phi \phi}^\epsilon+\delta \Gamma_{t \epsilon t} \bar{\Gamma}_{\phi \phi}^\epsilon \\
-\bar{\Gamma}_{t \epsilon \phi} \delta \Gamma_{\phi t}^\epsilon-\delta \Gamma_{t \epsilon \phi} \bar{\Gamma}_{\phi t}^\epsilon
\end{gathered}
\end{equation}
\begin{equation}
\begin{gathered}
\delta R_{r \theta r \theta}=\bar{D}_r \delta \Gamma_{r \theta \theta}-\bar{D}_\theta \delta \Gamma_{r \theta r} \\
+\bar{\Gamma}_{r \epsilon r} \delta \Gamma_{\theta \theta}^\epsilon+\delta \Gamma_{r \epsilon r} \bar{\Gamma}_{\theta \theta}^\epsilon \\
-\bar{\Gamma}_{r \epsilon \theta} \delta \Gamma_{\theta r}^\epsilon-\delta \Gamma_{r \epsilon \theta} \bar{\Gamma}_{\theta r}^\epsilon
\end{gathered}
\end{equation}
and 
\begin{equation}
\begin{gathered}
\delta R_{r \phi r \phi}=\bar{D}_r \delta \Gamma_{r \phi \phi}-\bar{D}_\phi \delta \Gamma_{r \phi r} \\
+\bar{\Gamma}_{r \epsilon r} \delta \Gamma_{\phi \phi}^\epsilon+\delta \Gamma_{r \epsilon r} \bar{\Gamma}_{\phi \phi}^\epsilon \\
-\bar{\Gamma}_{r \epsilon \phi} \delta \Gamma_{\phi r}^\epsilon-\delta \Gamma_{r \epsilon \phi} \bar{\Gamma}_{\phi r}^\epsilon 
\end{gathered}
\end{equation}
Therefore, we obtain 
\begin{equation}
\begin{aligned}
\Re\left(\Psi_4\right)=-\frac{1}{2}[ & \bar{D}_t \delta \Gamma_{t \theta \theta}-\bar{D}_\theta \delta \Gamma_{t \theta t} \\
& -\bar{D}_t \delta \Gamma_{t \phi \phi}+\bar{D}_\phi \delta \Gamma_{t \phi t} \\
& -\bar{D}_r \delta \Gamma_{r \theta \theta}+\bar{D}_\theta \delta \Gamma_{r \theta r} \\
& \left.+\bar{D}_r \delta \Gamma_{r \phi \phi}-\bar{D}_\phi \delta \Gamma_{r \phi r}\right] \\
&-\frac{1}{2}\left[Q_{t \theta t \theta}-Q_{t \phi t \phi}\right. & \\
& \left.-Q_{r \theta r \theta}+Q_{r \phi r \phi}\right]
\end{aligned}
\label{real psi 4}
\end{equation}
where 
\begin{equation}
Q_{\alpha \beta \gamma \delta}=\bar{\Gamma}_{\alpha \epsilon \gamma} \delta \Gamma_{\beta \delta}^\epsilon+\delta \Gamma_{\alpha \epsilon \gamma} \bar{\Gamma}_{\beta \delta}^\epsilon-\bar{\Gamma}_{\alpha \epsilon \delta} \delta \Gamma_{\beta \gamma}^\epsilon-\delta \Gamma_{\alpha \epsilon \delta} \bar{\Gamma}_{\beta \gamma}^\epsilon
\end{equation}
Similarly, for the imaginary part we have 
\begin{equation}
\delta R_{t \theta t \phi}=\bar{D}_t \delta \Gamma_{t \theta \phi}-\bar{D}_\phi \delta \Gamma_{t \theta t}+Q_{t \theta t \phi}
\end{equation}
and 
\begin{equation}
\delta R_{r \theta r \phi}=\bar{D}_r \delta \Gamma_{r \theta \phi}-\bar{D}_\phi \delta \Gamma_{r \theta r}+Q_{r \theta r \phi} 
\end{equation}
Thus,
\begin{equation}
\begin{aligned}
\Im\left(\Psi_4\right)= & \bar{D}_t \delta \Gamma_{t \theta \phi}-\bar{D}_\phi \delta \Gamma_{t \theta t} \\
& -\bar{D}_r \delta \Gamma_{r \theta \phi}+\bar{D}_\phi \delta \Gamma_{r \theta r} \\
& +Q_{t \theta t \phi}-Q_{r \theta r \phi} 
\end{aligned}
\label{imag psi 4}
\end{equation}

Finally, in the linearised regime the $Q_{\alpha \beta \gamma \delta}$'s disappear and we obtain
\begin{equation}
\begin{aligned}
\Psi_4= & -\frac{1}{2}\left[D_t \delta \Gamma_{t \theta \theta}-D_\theta \delta \Gamma_{t \theta t}-D_t \delta \Gamma_{t \phi \phi}+D_\phi \delta \Gamma_{t \phi t}\right. \\
& \left.-D_r \delta \Gamma_{r \theta \theta}+D_\theta \delta \Gamma_{r \theta r}+D_r \delta \Gamma_{r \phi \phi}-D_\phi \delta \Gamma_{r \phi r}\right] \\
& +i\left[D_t \delta \Gamma_{t \theta \phi}-D_\phi \delta \Gamma_{t \theta t}-D_r \delta \Gamma_{r \theta \phi}+D_\phi \delta \Gamma_{r \theta r}\right] 
\end{aligned}
\label{linearised psi}
\end{equation}
Notice that for a strictly spherically symmetric reduction, there are no gravitational waves, so $\Psi_4=0$. To construct a nonzero $\Psi_4$ we must introduce perturbations that break spherical symmetry. From argument 1, we know that perturbing $K_T$ and $n_r$ can be related to the Regge-Wheeler function $\Phi$. 

In the following part of this argument, we substitute the Ricci rotation coefficients present in the toy model into eqs.(\ref{real psi 4},\ref{imag psi 4}), relaxing the spherical symmetry assumption, i.e. the Ricci rotation coefficients not present in the toy model are not automatically set to zero. In this way, we can check for interactions with the Ricci rotation coefficients present in the toy model.\\

Recall in WEBB we have 
\begin{equation}
\begin{gathered}
a_a=\Gamma_{a 00} \\
K_{a b}=\Gamma_{b 0 a} \\
N_{a b}=\frac{1}{2} \epsilon_b^{c d} \Gamma_{c d a} 
\end{gathered}
\end{equation}
Now we express the related perturbative connection $\delta \Gamma_{\alpha \beta \gamma}$. The time-space components become 
\begin{equation}
\delta \Gamma_{i 0 j}=\delta K_{j i}
\end{equation}
and 
\begin{equation}
\delta \Gamma_{i j 0}=\epsilon_{i j}{ }^k \delta \omega_k
\end{equation}
but for a Fermi-transported tetrad these disappear since $\omega_k=0$. For purely spatial indices, we have 
\begin{equation}
\delta \Gamma_{i j k}=\epsilon_{i j}^l \delta N_{l k}
\end{equation}
Hence, for the WEBB variables with time-space components we have 
\begin{equation}
\begin{aligned}
\delta \Gamma_{t \theta \theta} & =-\delta K_{\theta \theta} \\
\delta \Gamma_{t \phi \phi} & =-\delta K_{\phi \phi} \\
\delta \Gamma_{t \theta \phi} & =-\delta K_{\phi \theta} \\
\delta \Gamma_{t \theta t} & =-\delta a_\theta \\
\delta \Gamma_{t \phi t} & =-\delta a_\phi
\end{aligned}
\end{equation}
and for purely spatial indices, since $\delta \Gamma_{r \theta \theta}=\epsilon_{r \theta}{ }^m \delta N_{m \theta}$ and $\epsilon_{r \theta \phi}=+1$, we have  
\begin{equation}
\begin{aligned}
\delta \Gamma_{r \theta \theta} & =\delta N_{\phi \theta}\\
\delta \Gamma_{r \phi \phi} & =-\delta N_{\theta \phi} \\
\delta \Gamma_{r \theta \phi} & =\delta N_{\phi \phi} \\
\delta \Gamma_{r \theta r} & =\delta N_{\phi r} \\
\delta \Gamma_{r \phi r} & =-\delta N_{\theta r}
\end{aligned}
\end{equation}
Substituting these into eq.\ref{linearised psi}, we obtain 
\begin{equation}
\begin{aligned}
\Re\left(\Psi_4\right)=-\frac{1}{2}[ & -D_t \delta K_{\theta \theta}+D_\theta \delta a_\theta \\
& +D_t \delta K_{\phi \phi}-D_\phi \delta a_\phi \\
& -D_r \delta N_{\phi \theta}+D_\theta \delta N_{\phi r} \\
& \left.-D_r \delta N_{\theta \phi}+D_\phi \delta N_{\theta r}\right] \\
& + \text { background } \times \delta(K, N, a, \omega) 
\end{aligned}
\end{equation}
and 
\begin{equation}
\begin{aligned}
\Im\left(\Psi_4\right)= & -D_t \delta K_{\phi \theta}+D_\phi \delta a_\theta \\
& -D_r \delta N_{\phi \phi}+D_\phi \delta N_{\phi r} \\
& + \text { background } \times \delta(K, N, a, \omega) 
\end{aligned}
\end{equation}
and rearranging we obtain
\begin{equation}
\begin{aligned}
\Psi_4= &\mathcal{D}_t\left[\left(\delta K_{\theta \theta}-\delta K_{\phi \phi}\right)+2 i \delta K_{\theta \phi}\right]\\
&+\mathcal{D}_r\left[\left(\delta N_{\phi \theta}+\delta N_{\theta \phi}\right)+2 i \delta N_{\phi \phi}\right]+\cdots
\end{aligned}
\label{final psi 4}
\end{equation}

from which we see that one polarization is controlled by 
$\delta K_{\theta \theta}-\delta K_{\phi \phi}
$ with the related $N$-sector pieces, while the other polarization is controlled by $\delta K_{\theta \phi}$ and the the related $N$-sector pieces, and the omitted terms are the background couplings coming from the $\bar{\Gamma} \delta \Gamma$ part of the linearized curvature, of the type $a_a \delta K_{b c}, \quad K_{a b} \delta K_{c d}, \quad N_{a b} \delta N_{c d}$.\\

In particular, focusing on the combination 
\begin{equation}
\left(\delta K_{\theta \theta}-\delta K_{\phi \phi}\right)+2 i \delta K_{\theta \phi}
\end{equation}
\label{K eq}
notice that $\left(\delta K_{\theta \theta}-\delta K_{\phi \phi}\right)$ is only equal to zero if the spherical symmetry of the background sets $\delta K_{\theta \theta}=\delta K_{\phi \phi}$. However, we established through argument 1 that a perturbation in $n_r$ and $K_T$ can be related to a non-zero Regge-Wheeler function of the $l=2$ sector. Hence, suppose that, relaxing the perfect spherical symmetry assumption if we were to evolve the full 3+1 dimensional system $\delta K_{\theta \theta} \neq \delta K_{\phi \phi}$ and $\delta K_{\theta \phi} \neq 0$ but unknown.

Then, observing the terms appearing in eq.\ref{final psi 4}, we see that while $K_T$ and  $n_r$ are scalars in the background, they can have a non-zero projection onto the spin-2 sector once perturbations break the spherical symmetry because of the background coupling terms $\bar{\Gamma} \delta \Gamma$. In fact, the omitted terms in eq.\ref{final psi 4} have structure like
\begin{equation}
\bar{K}_T \delta K_{\theta \phi}, \quad \bar{n}_r \delta K_{\theta \phi}, \quad \bar{a}_r \delta N_{\theta \phi} 
\end{equation}
and evolution equations will generically contain terms like:
\begin{equation}
\partial_t \delta K_{\theta \phi} \sim \bar{n}_r \cdot(\text { derivative of scalar perturbations })
\end{equation}
For these reasons, we propose a function of the form 
\begin{equation}
\Psi_4 \propto \partial_t \delta K_T-i \partial_t \delta n_r
\end{equation}
as a suitable approximation of the Regge-Wheeler function projected onto the variables that are available in the toy model we considered that is expected to contain the $l=2$ QNMs frequencies. 

We use initial gaussian perturbations of the form 
\begin{equation}
\delta K(t  =0, r)=-A \exp \left[-\frac{\left(r-r_0\right)^2}{\sigma^2}\right] 
\label{perturb}
\end{equation}
as shown in fig.\ref{init pert _1}

\end{document}